\documentclass[useAMS,usenatbib]{mn2e}

\usepackage{graphicx}
\usepackage{float}

\title[Massive planets on inclined orbits]{Interaction between massive planets on inclined orbits  and circumstellar discs}
\author[M. Xiang-Gruess  and  J. C. B. Papaloizou]{M. Xiang-Gruess$^{1}$  \thanks{E-mail:
mx216@cam.ac.uk } and  J. C. B. Papaloizou$^{1}$
 \\
$^{1}$ Department of Applied Mathematics and Theoretical Physics, University of Cambridge, Wilberforce Road, \\
 Cambridge CB3 0WA, United Kingdom
}
\begin{document}

\date{Accepted . Received ; }

\pagerange{\pageref{firstpage}--\pageref{lastpage}} \pubyear{2012}

\maketitle

\label{firstpage}

\begin{abstract}
We study the interaction between  massive planets  and a gas disc with a mass in the range expected for protoplanetary discs.
We use SPH simulations to study the  orbital  evolution  of a massive planet as well as  the dynamical  response of the 
disc for planet masses between 1 and $6\ \rmn{M_J}$ and  the full range
of initial relative orbital  inclinations.

We  find that gap formation can occur for planets in inclined orbits as well as for coplanar orbits as expected.
For given planet mass, a threshold relative  orbital inclination exists under which a gap  forms.
This threshold increases with planet mass.
Orbital migration manifest through  a decreasing semi-major axis is seen in all cases.

At  high relative inclinations,
the inclination decay rate increases for increasing planet mass and decreasing initial relative inclination
as is expected from estimates based on dynamical friction between planet and disc.
 For an  initial semi-major axis  of $5$~AU and  relative
inclination of $i_0=80^\circ,$
the times required for the inclination
to  decay  by  $10^\circ$ is $\sim10^{6}\ \rmn{yr}$ and $\sim10^{5}\ \rmn{yr}$  for $1\ \rmn{M_J}$ and $6\ \rmn{M_J}$
respectively, these times scaling in the usual way for larger  initial orbits.
For retrograde planets, the inclination always evolves towards  coplanarity with the disc, with the rate of evolution being fastest
for orbits with  $i_0 \to 180^\circ.$ The indication is thus  that, without taking account of subsequent operation of phenomena such as the Lidov-Kozai effect,  planets with mass $~1\ \rmn{M_J}$ initiated in circular orbits
with semi-major axis $\sim 5$~AU and   $i_0 \sim 90^\circ$ might only just  become coplanar, 
as a result of frictional effects,  within the disc lifetime. In other cases highly inclined orbits will survive only
if they are formed after the disc has mostly dispersed.

Planets on inclined orbits  warp the disc by an extent that is negligible for
$1\ \rmn{M_J}$ but increases with increasing mass becoming quite significant for a planet of mass  $6\ \rmn{M_J}$.
In that  case, the disc can gain a total inclination of up to $15^\circ$
together with  a warped inner structure with  an inclination of up to $\sim 20^\circ$ relative to  the outer part.
We also find a  solid body precession of both the total disc angular momentum vector
and the planet orbital momentum vector about the total angular momentum vector,
with the angular velocity of  precession  decreasing  with increasing relative inclination as expected in that case.

Our results illustrate that the influence of an inclined massive planet on a
protoplanetary disc  can lead to significant changes of the disc structure and orientation which can  in turn
 affect the orbital evolution of the planet significantly.  A three-dimensional treatment of the disc is then essential
 in order to capture all relevant dynamical  processes in the composite system.

\end{abstract}


\section{Introduction}

In the past few decades, the number of detected extrasolar planets around other main sequence stars has increased dramatically,
so that  at the time of writing,  more than 840 planets have been discovered.
Well studied  planet formation  scenarios  such as  core accretion  \citep{Miz1980, Pol1996} or
disc fragmentation \citep{May2002} involve the planet forming in a disc with the natural expectation
that the orbit
 should be  coplanar. However, Rossiter McLaughlin measurements  of close orbiting hot Jupiters
indicate that around $40\%$ have angular momentum vectors  significantly misaligned with  the
angular velocity vector of the central star \citep[e.g.][]{Tri2010, Alb2012}. As the stellar and disc angular velocities
are naturally expected to be aligned, this would imply an inclination of the planet orbit relative to the nascent protoplanetary disc.  But note that  \citep[e.g.][]{Lai2011}
have proposed that a magnetic interaction between the central star and protoplanetary
disc could lead to the stellar angular momentum vector
 being misaligned with that of the disc, in  which case misaligned planetary orbits
need not be inferred.  However, 
\citet{ Wat2011} 
have compared stellar rotation axis inclination angles with measured debri-disc inclinations
and found no evidence for 
 misalignment.

Several  scenarios have been proposed to explain the origin of  misalignments between the planetary orbital angular  momentum vector
 and the stellar rotation  axis for close in planets.
The first involves excitation  of very high eccentricities, either through  the Lidov-Kozai
effect induced by the  interaction with a distant companion \citep[e.g.][]{Fab2007, Wu2007},
or through planet-planet scattering or chaotic interactions  
\citep[e.g.][]{Weid1996,Rasio1996,Pap2001,Nag2008}.
This is is then followed by orbital circularization due to interaction with the central star, although aspects
of this process are not yet understood in detail \citep{Daw2012}.
It is possible that  high eccentricities and inclinations  for massive planets are generated when the disc is present.
It is then important, as has been noted by \citet{Daw2012},
 to understand the  interactions with the disc and  consequent  time scales for realignment.

Another method for producing high inclinations
has been indicated by \cite{Tho2003}, who  have studied the evolution of two giant planets in resonance,
with an approximate analytical expression for the influence of a gas disc in producing  orbital migration.
Their calculations suggest that resonant inclination excitation
can occur when through the interaction between the planets the eccentricity of the inner planet
reaches a threshold value of $e_1 \geq 0.6$. Inclinations gained by the resonant pair of planets
can reach values up to $\sim 60^\circ$. The effects of disc planet interactions may be significant here also.

Another explanation for the origin of  misaligned planets is connected with the possibility that the orientation of the disc
changes, possibly due to lack of alignment of the source material or due to gravitational encounters
with passing stars,  either before or after the planet forms \citep[e.g.][]{Bat2010,Thi2011}.
However, up to now, there is no observational evidence for such disc  misalignment \citep{Wat2011}.

While the formation of  planets on misaligned orbits  has been a frequent topic of debate, 
the evolution of  massive planets  on  misaligned orbits interacting with a disc is still not well understood.
Many  hydrodynamical simulations have been  performed in order  to study the influence of a protoplanetary
disc on the evolution of planets. The majority of these simulations have  considered
planets that are coplanar with the disc and been two dimensional and  used grid-based methods
\citep[see][for a review and references therein]{Pap2007,Kle2012}.

Planets with initial orbital inclinations with respect to a disc
have been studied  \citep[e.g.][]{Cre2007, Bit2011}. In these works the evolution of
planets with masses  up to a maximum of  $1\rmn{M_J}$ with different initial eccentricities and relatively small
 initial inclinations up to a maximum of $15^\circ$ interacting with three-dimensional isothermal and radiative discs are considered.
They showed damping for both eccentricity and inclination with the planets circularising  in the disc after a few hundred orbits.

In this paper we are interested in extending studies of disc planet interactions involving  planets with orbital
planes misaligned with that of the  disc
to consider larger planet masses of up to several Jupiter masses and the full range of inclinations.
In such cases the interaction with the disc is more likely to be described by dynamical friction
than being the result of the application of resonant torques that is applicable in the coplanar case
\citep[see for example][]{Pap2000, Rei2012, Tey2013}.
Numerical simulations of warped discs in close binary systems were performed by e.g.
\cite{Lar1996, Fra2010}.
These works indicate  that thick discs with low viscosities have efficient warp communication 
which allows them to precess as rigid bodies with little warping or twisting while thinner discs 
can develop twists before reaching a state of rigid-body precession.
Possible warping of a disc by giant planets has yet to be studied in detail.
It has been estimated from linearized calculations \citep{Tey2013} that
disc warping should be a small effect for sub Jovian mass planets, this being confirmed by the results
of this paper. However, it turns out  to be significant for larger
planet masses that can produce significant nonlinear  effects, therefore we study these effects  in this paper.

It has been shown that Smoothed particle hydrodynamics (SPH) simulations are capable   of being applied to the problem of planet-disc interaction
\citep[e.g.][]{Sch2004,Val2006} though in comparison to grid-based simulation methods, they incur greater computational expense.
But in contrast to grid-based methods, as they adopt a Lagrangian approach,
SPH simulations can be readily used to simulate a gas disc  with a free boundary in three dimensions
with the disc having the freedom to change its shape at will, thus we adopt this approach  here. 
Similarly, \cite{Mar2009} studied the interaction between a Jupiter mass planet and a disc using the SPH method.
Initial eccentricities ranging from 0 to 0.4 and an initial inclination of $20^\circ$ were considered.
The disc is simulated by the SPH code VINE that is similar to GADGET-2 and a locally isothermal equation of state. 
They find that  both the eccentricity and the inclination are damped on a timescale much lower that the disc lifetime ($\sim 10^6$ yr).

In this paper we consider a range of planet masses and the full range of orbital inclinations.
In addition to studying the warping response of the disc, we estimate the timescales for orbital evolution 
and attainment of coplanarity. We also make a preliminary study of the potential exchange of inclination and eccentricity
that might be expected for high inclination orbits through the operation of  an adaption of the  Lidov-Kozai mechanism.
The plan of the paper is as follows:

In Section \ref{sec:sim_details}, we describe our simulation code with   
a description of the modified locally isothermal equation of state we adopted (Section \ref{sec:eos}).
Some details concerning the smoothing length and artificial viscosity
are given in Section \ref{sec:artv}.
In Section \ref{sec:IC}, we describe the general setup for  the disc,  planet and the central star.
In Section \ref{sec:theo}, we give a brief  overview of the interaction between a massive planet and disc. 
We give an  analytic estimate of the evolution time scale for  a planet interacting with  a disc 
based on considerations of dynamical friction, in order to compare with our simulations, 
in Section \ref{sec:estim}. In addition, we consider the possibility of the Lidov-Kozai effect (Section \ref{sec:LidovK}). 
In Section \ref{sec:coplanar}, we discuss numerical results for  a massive planet on a  coplanar circular orbit. 
We study the evolution of  planets initiated on inclined orbits  in Section \ref{sec:inclined}
considering the  evolution of both initially prograde and  retrograde circular orbits.
In Section \ref{sec:response}, we study disc warping  and precession.
In Section \ref{sec:incecc}, we study a planet starting on an eccentric inclined orbit
finding evidence for some exchange of inclination and eccentricity as might be expected from 
a Lidov-Kozai like process.
The behaviour of an inclined planet interacting with discs of different masses is considered in Section \ref{sec:M_d}.
Finally the results are  summarised and discussed in Section \ref{sec:concl}.

\section{Simulation details} \label{sec:sim_details}
We  have performed simulations using a modified version of the publically available code GADGET-2 \citep{Spr2005}.
GADGET-2 is a hybrid N-body/SPH code capable of modelling
both fluid and distinct  fixed or  orbiting  massive bodies.
In our case the central star, of mass $M_*,$  is fixed and the planet, of mass $M_p,$  orbits as a distinct massive body. 
We adopt spherical polar coordinates $(r, \theta,\phi)$
with origin at the centre of mass of the central star.

\noindent  The gaseous disc is represented by SPH particles. 
An  important issue  in N-body/SPH simulations is the choice 
of the gravitational softening lengths.
Ideally, gravitational force computations without  softening 
 should be used to determine  the gravitational interactions between particles.
However, for numerical reasons, a softening length has to be introduced.
 This is because  SPH particles would experience very high  gravitational accelerations in the  vicinity of massive objects.
 In turn, this would result in unacceptably small time steps that prevent simulations from effectively proceeding.
For this reason, we adopt  gravitational softening lengths that are sufficiently large
 to enable the  calculation to proceed, and in the case of the planet,  smaller than physically relevant length  scales such as the  disc  height $H$ 
 and the  planetary Hill radius. It can be associated with a physical size of material considered to be bound to the planet.
 Here we study planets of masses $M_p\geq 1\ \rmn{M_J}$ with an orbital length scale of  $R_p=$~5~AU and discs with  aspect ratio $H/r=0.05.$
Thus  the Hill radius defined as $R_H=R_p \left(M_p/(3 M_*) \right)^{1/3}$ is 0.35 AU for a Jupitermass planet at 5 AU.
 At $r=5\ \rmn{AU}$, $H=0.25\ \rmn{AU}$.
In our simulations, the central  star and the planet are unsoftened when interacting with each other
 allowing them to undergo close encounters accurately, {although they do not actually occur in our simulations.}
For the  computation  of the gravitational interaction between the
massive bodies  and the SPH particles, we adopted fixed  softening lengths  
$\varepsilon_*=0.4\ \rmn{AU}$ and $\varepsilon_{p}=0.1\ \rmn{AU}$ for the central star and the planet, respectively.
The former value controls conditions near the central star. These are  not significant for the dynamics of interest in our simulations
which only depend on the value of the central stellar mass.
 The latter value  $\varepsilon_{p}$ controls conditions close to the planet and is significantly 
 smaller than both $R_H(1\ \rmn{M_J}, 5\ \rmn{AU})$ and $H(5\ \rmn{AU})$.

The star and the planet 
are allowed to  accrete gas particles that approach them very closely.
The reason for doing this is to prevent the possibility of strong unresolved forces occuring close to massive objects.
In this respect it works in tandem with softening.
  In order to implement accretion, we followed  the procedure  of \cite{Bat1995} for so-called "sink particles". 
This was applied such that for the star and planet, the outer accretion radii were fixed during the simulation to be  
$R_{accr,*}=7\times 10^{11}\ \rmn{cm}$ and $R_{accr, p}=7\times 10^{10}\ \rmn{cm}$ respectively.
SPH particles that come within the accretion radius of a sink particle, are accreted if they fullfill all conditions for accretion. 
Firstly, the particle must be bound to the sink particle. Secondly, its specific angular momentum about the sink particle must be less than that required for it to form a circular orbit at $R_{acc}$ about the sink particle.
These conditions ensure that particles which would normally leave the accretion radius are not accreted.
Thirdly, the particle must be more tightly bound to the candidate sink particle than to any other sink particle.
In our case  this is to allow for the possibility that, given the very different masses of the star and planet, 
 a SPH particle could  pass  through the accretion radius of the planet and 
 yet be  accreted by the central star. 

An inner accretion radius can also be defined such that   SPH particles are accreted regardless of the tests \citep{Bat1995}.
By using an inner accretion radius,  particles are prevented from being accelerated by very close encounters with the sink particle
that might otherwise dramatically slow down the simulation.
\cite{Bat1995} suggests that the inner accretion radius should be 10-100 times smaller than the outer accretion radius. 
Due to our very small outer accretion radii, the choice of the inner accretion radii to be 0.5 $R_{acc}$ was found to be adequate for this purpose. 

In practice, for the equations of state  and softening parameters we used for most
 of our simulations, because of the relatively small accretion radii adopted,  the
accretion of gas particles was found to play only  a minor role, producing negligible  changes
to the masses of the planet and star (see also the discussion in section \ref{5.1}).

\subsection{Equation of state} \label{sec:eos}
For the hydrodynamical simulations, we applied a locally isothermal equation of state 
with the modification described by \cite{Pep2008}.
The reason for modification is because, as \citet{Pep2008} have pointed out,
if the temperature  characteristic of the main disc is adopted in the vicinity of the planets,
 a high concentration of gas  leading to rapid accretion develops in the close vicinity of massive planets.
Accordingly, we adopt a sound speed given by  \citep{Pep2008}
\begin{eqnarray}
 c_s=\frac{h_s r_s h_p r_p}{\left[ (h_sr_s)^n + (h_p r_p)^n \right]^{1/n}} \sqrt{\Omega_s^2 + \Omega_p^2}\ , \label{eq:Pep_mod}
\end{eqnarray}
Here $r_s= |{\bf r}-{\bf r}_s|$ and  $r_p=|{\bf r}-{\bf r}_p|$ are
 the distances to the central star and the planet respectively, and
$\Omega_s$ and $\Omega_p$ are the angular velocities in the circumstellar and circumplanetary discs.
For pure Keplerian motion they are given by
\begin{equation}
 \Omega_s=\sqrt{\frac{\rmn{G} M_*}{r_s^3}}\hspace{2mm}    {\rm and} \hspace{2mm}  
 \Omega_p=\sqrt{\frac{\rmn{G} M_p}{r_p^3}}\hspace{2mm} {\rm respectively.}
\end{equation}
The disc aspect ratio is $h_s=H/r_s$  with  $H$ being  the disc scale height. The parameter $n$ 
was taken to be  $n=3.5$. 
For the simulations we take $H/r_s = c_s/v_\varphi=0.05$, where  $v_\varphi =r_s \Omega_s$
 is the rotational velocity. 
As $r_p \rightarrow 0,$  the sound speed in the circumplanetary disc is $h_pr_p\Omega_p$
and accordingly $h_p$ is the aspect ratio in that limit.  This should be chosen large enough
so that pressure forces prevent excessive gas accumulation on the planet. At the same time, it must be chosen large enough to prevent values of the  sound  speed that are smaller than the unmodified value, 
 at locations that are a moderate distance away from the planet.
 \citet{Pep2008} suggest $h_p\geq 0.4$. 
A discussion  of  the effects resulting from different choices of  $h_p$ can be found in
 appendix \ref{ap:peplinski}. For our simulations, we adopted  $h_p=0.6$ which was large enough to
 prevent unrealistically fast gas accretion while not causing the sound speed to become smaller than the
 unmodified value in all cases.

\subsection{Smoothing length and artificial viscosity} \label{sec:artv}

For our  SPH calculations, the smoothing length was adjusted so that the number of  nearest  neighbours  to any particle contained within a sphere of radius equal to the local smoothing length was  $40 \pm 5$. 
The pressure is given by  $p=\rho c_s^2$. 
Thus apart from  the vicinity of the planet, the temperature in the disc is $\propto r^{-1}.$ 
The artificial viscosity parameter $\alpha$ of GADGET-2 \citep[see equations (9) and (14) of][]{Spr2005} 
was taken to be $\alpha =0.5$. 

The artificial viscosity is modified by the application of a viscosity-limiter to reduce artificially induced  angular
momentum transport in the presence of shear flows. This is especially important  for the study of Keplerian discs.
In order to calibrate the diffusive effects resulting from this viscosity, we applied the ring spreading test to determine an effective kinematic viscosity measured through the $\alpha_{SS}$ parameter \citep{Sha1973}. 
Details are given in appendix \ref{ap:ring_test}.
A value of $\alpha= 0.5,$ showed the best  match to the analytic ring spreading solution with  $\alpha_{SS}=0.02$
and was hence used in all simulations.
SPH simulations accordingly model a viscous disc, with a viscosity that behaves like a conventional Navier Stokes viscosity
in this context. It should be noted that the effective viscosity in  protoplanetary discs is likely to involve magnetic fields
and behave differently in some contexts. However many of the simulations undertaken here involve impulsive high velocity interactions
between planet and disc that are characteristically described by dynamical friction (see below) and for which the disc viscosity is
not expected to play a major role. The remaining simulations involve coplanar, inwardly migrating, gap forming planets for which
the interaction with the disc is nonlinear and long range. The characteristic behaviour again is not expected to be sensitive
to details of the viscosity \citep[e. g.][]{Nel2003, Bar2011}.

\section[]{Initial conditions} \label{sec:IC}
We study a system composed of a central star of one solar mass $\rmn{M_\odot}$, a gaseous disc and a massive planet.
The disc is set up such that the angular momentum vector for all particles 
was in the same direction enabling a mid plane for the disc to be defined.
Planet masses  in the range 1 - $6\ \rmn{M_J}$  were initiated   in circular orbits with inclinations 
in the range $i_0=[0;80]^\circ$  with respect to the initial disc mid plane. The semi-major axis of the planet was set to $a = 5\ \rmn{AU}$.
The  particle distribution was chosen
to  model  a  surface density profile  such that
\begin{eqnarray}
\Sigma=\Sigma_0 R^{-1/2}. 
\end{eqnarray}
Here $\Sigma_0$ is a constant and $R$ is the radial coordinate of a point in the midplane, here
taken to  occupy  the radial domain $[0, R_{out}-\delta]$, with  $R_{out}= 5a$ and $\delta =0.4 a$.
  A taper was applied  such that the surface density was set to decrease linearly to zero
  for $R$ in the  interval $[ R_{out}-\delta,  R_{out}+\delta].$
The total disc radial domain is hence $[0, R_{out}+\delta]$. 
 
\noindent  The disc mass is given by
\begin{eqnarray}
M_D=2\pi \int_{0}^{R_{out}} \Sigma(r)r dr=\frac{4}{3} \pi \Sigma_0 R_{out}^{3/2}\ r,
\end{eqnarray}
which is used to determine $\Sigma_0.$
The disc particles were set up in a state of pure Keplerian rotation velocity according to
\begin{eqnarray}
 v_\varphi=\sqrt{r_s\frac{d\Phi_*}{dr_s}}\ .
\end{eqnarray}
The  radial velocities were set to be zero.
Here $\Phi_*$ is the gravitational potential due to the central star.
In the innermost region around the central star, the disc properties (e.g. $\Omega_s$ and accordingly $c_s$) are modified by  the  softening.

In this paper, only  discs for which self-gravity can be neglected  are studied. For this to be possible with  a locally 
isothermal equation of state, the Toomre parameter has to fulfil the requirement 
\begin{eqnarray}
 Q=\frac{\Omega c_s}{\pi G \Sigma} > 1.4\ .
\end{eqnarray}
For the case $H/r_s=0.05$, the Toomre parameter can be expressed as
\begin{eqnarray}
Q=\frac{0.05 M_*}{r_s^2 \pi  \Sigma }
=\frac{ R_{out}^{3/2}  M_*}{15 M_D} r_s^{-3/2}\ . \label{eq:Q,r^-1/2}
\end{eqnarray}
For a disc with any arbitrary outer radius $R_{out}$, the maximum allowed disc mass $M_{D,max},$ 
which  fulfils the requirement that  $Q(R_{out})=1.4$ at its outer edge is 
\begin{eqnarray}
M_{D, max}=\frac{M_*}{15 Q(R_{out})}
=\frac{M_*}{21}.
\end{eqnarray}
Hence, for any arbitrarily chosen outer radius $R_{out}$, the total disc mass has to be set smaller than $M_*/21$ $\sim 0.05 M_*$. For simulations shown in this paper, the disc mass is $M_D=0.01\ M_\odot$  unless stated otherwise which corresponds to
\begin{eqnarray}
 \Sigma_0=\frac{3 M_D}{4 \pi R_{out}^{3/2}}=\frac{0.03\ \rmn{M_\odot}}{20\sqrt{5} \pi a^{3/2}}\ .
\end{eqnarray}
The disc mass surface density at the location of a planet with  $a=5\ \rmn{AU}$ is
   then $76
\ \rmn{g\ cm^{-2}}$.

In addition to self-gravity, energy transfer and 
mass infall onto the disc are neglected.
At the start of a simulation, the disc is allowed to evolve for 30 orbits before the planet is introduced.
By doing this, we minimise the effects of  initial transients in the disc arising from the setup.

In order to examine the effect of changing the number of SPH
particles, we performed resolution studies with both coplanar  planets and   planets
initiated on inclined orbits. These are described in appendix \ref{ap:resolution}.
The  number of SPH particles chosen for most of  our simulations was 2$\times 10^5$. 
This enabled enough  simulations to be
 run for $\sim$ 500 orbits or $500 \times P(\mathrm{5\ AU})=500 \times 11.18\ \rmn{yr}=5600 \ \rmn{yr}$
that a reasonable exploration of phase space could be made while retaining adequate accuracy.
Finally we note that although for definiteness we have fixed length and time units to correspond to 5~AU,
they can be scaled to other values with the usual  corresponding scaling of the units of length and  time.

\section{Orbital evolution of a single massive planet} \label{sec:theo}

\subsection{General overview}

The evolution of a system composed of a central star, a planet and a gaseous disc is dependent on the
relative masses of the different components. 
Low-mass planets produce relatively small amplitude  density  perturbations 
such that the global disc evolution is not significantly  modified.
In contrast, a massive planet can influence the global evolution of a disc significantly in two ways.
Firstly, material may be cleared from its neighbourhood by tidal interaction, creating a deep and wide gap.
Secondly, it is expected that massive planets on inclined orbits could  modify the three-dimensional structure of the disc
by producing large scale twists and warps. However, this has not yet been studied in detail.


In  the mass domain   $M_p \in  [1 \rmn{M_J} ;6 \rmn{M_J}]$
considered here, a planet in a coplanar orbit is expected  and been found  to open a gap in the disc. 
\citep[e.g.][]{Lin1979, Var2004, Cri2006}.
In addition net torques acting on the planet  result in planet migration. 
When a  planet opens a gap in its host disc, provided it is not too massive, 
it migrates inward on the  viscous timescale of the disc (type II migration).

When the orbit of the planet is significantly inclined with 
respect to the disc mid plane, the interaction with the disc is expected to differ from the  coplanar case.
Impulsive interactions  occuring twice per orbit that can be described through an
approach based on dynamical friction are expected.
Recently, \cite{Rei2012} made a local study of low-mass planets on highly inclined orbits 
and was able to confirm the dynamical friction viewpoint
and show that the time-scale for realignment in such cases
was long compared to the disc lifetime meaning that such inclined orbits should survive the presence of a disc.  
For massive planets, the interaction is expected to be stronger with  the evolutionary time scales
correspondingly reduced.  In addition, the planet is more likely to  affect the  structure and evolution of the disc.

\subsection{Estimate of time scale for orbital evolution} \label{sec:estim}
Here we estimate an approximate time scale for orbital evolution  assuming that
the interaction between the planet and the disc can be described as occurring through dynamical friction.
It has been noted by a number of authors \citep[e.g.][]{Pap2000, Rei2012, Tey2013}
that planets on orbits with high eccentricity and/or high inclination  pass through the disc with a high supersonic relative velocity 
such that the interaction can be approximated as being with pressure-less ballistic fluid particles.  
A dynamical friction calculation is then appropriate.  The frictional force per unit mass is then  \citep[e.g.][]{Rud1971, Rep1980, Ost1999} 
\begin{equation}
{\bf f}_D = - \frac{ 4\pi G^2 M_p\rho{\bf v}_{rel}   }{|{\bf v}_{rel}|^3  }\ln(r_1/r_2),\label{dynf}
\end{equation}
where ${\bf v}_{rel} $ is the relative velocity between the planet and disc gas of density $\rho.$
 Here $r_1$ and $r_2$ represent upper amd lower cut off length scales outside of  which
the analysis leading to (\ref{dynf}) that assumes a homogeneous medium becomes invalid.
There is some uncertainty in estimating these that in turn leads to an uncertainty in the force
of a factor of a few, and so we can only use (\ref{dynf}) to make rough estimates.
We consider a high inclination orbit and a thin disc such that the interaction may be viewed
as impulsive and occurring twice per orbit.
Each impulse produces a  small velocity change given by
\begin{equation}
\Delta {\bf v}= - \frac{ 4\pi G^2 M_p \ln(r_1/r_2)}{\sin i}    \int   \frac{\rho{\bf v}_{rel}   }{|{\bf v}_{rel}|^3 |{\bf v}| }dz ,
\end{equation}
where the disc mid plane is assumed to lie in a Cartesian $(x,y)$ plane, the inclination of the orbit to the disc is $i,$
the orbital speed is $ |{\bf v}|,$
and the integration is taken through the vertical extent of the disc.
For a thin disc this becomes
\begin{equation}
\Delta {\bf v}= - \frac{ 4\pi G^2 M_p \ln(r_1/r_2)\Sigma {\bf v}_{rel}   }{|{\bf v}_{rel}|^3 |{\bf v}|\sin i }, 
\end{equation}
where $\Sigma$ is the disc surface density.
This in turn leads to a characteristic time for the orbit to change measured in units
of the orbital period given by
\begin{equation}
T_D= |{\bf v}_{rel}| /(2|\Delta {\bf v}|) = \frac{|{\bf v}|^4 \sin i \sin^3(i/2)}{ \pi G^2 M_p \ln(r_1/r_2)\Sigma }, 
\end{equation}
where in order to obtain estimates, we have replaced $|{\bf v}_{rel}| $  by $2|{\bf v}|\sin(i/2).$
To proceed further we set  $r=a,$ $|{\bf v}|^4= G^2M_*^2/a^2$ and 
$\Sigma =3 M_D(a/R_{out})^{3/2}/(4\pi a^2).$
Thus,  we obtain
\begin{equation}
T_D =  \frac{M_*^2 \sin i\sin^3(i/2) }{\pi \Sigma  M_p a^2  \ln(r_1/r_2)}\equiv 
\frac{4M_*^2 \sin i\sin^3(i/2) }{ 3 M_p M_D  \ln(r_1/r_2)} \left(\frac{R_{out}}{a}\right ) ^{3/2} \label{TD}
\end{equation}
In order to  estimate $r_1,$ we note that when the planet is in the disc, the vertical scale should be limited by $H,$
but that $r_1$ should be some what larger to account for more distant material near the disc midplane. Accordingly
we take $r_1=2H.$ For $r_2$ we take the softening length which is around $0.4H$ at   $a=5\ \rmn{AU}.$
This is an  estimate of the scale below which material should be regarded as being bound to the planet.
Taking $M_D/M_* =0.01,$ $M_p/M_*= 0.001$, $ r_1/r_2=5,$ $R_{out}/a=5$ and 
$i=\pi/4,$ we obtain $T_D = 2\times 10^4 $ orbits, corresponding to
$4.3\times 10^5$~yr for $a=5\ \rmn{AU}.$

\begin{figure}
\centering
\includegraphics[width=5cm]{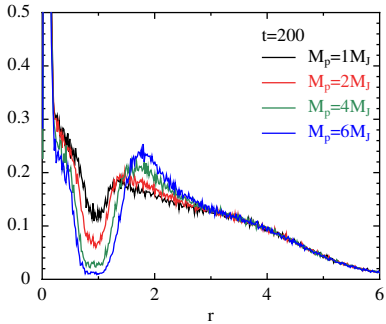}
\caption{ The azimuthally averaged  disc surface density profile (in units of  $\rmn{M_J/(5\ AU)^2}$) as a function of distance (in units of $\rmn{5\ \rmn{AU}}$) after $200$ hundred orbits for  planets  with $M_{p}=1, 2, 4,\hspace{1mm}  {\rm and}  \hspace{1mm}  6~ \rmn{M_J}$.
 The curves plotted are such that the density minimum in the centre of the gap decreases with
increasing planet mass. [All gap profiles were created using SPLASH \citep[][]{Pri2007}]} 
\label{fig:gap_coplanar}
\end{figure}

\subsection{Lidov-Kozai effect}\label{sec:LidovK}
In addition to dynamical friction, another mechanism that can affect the orbital evolution of planets
on orbits that are highly inclined to the disc is the Lidov-Kozai effect \citep{Koz1962}.
This causes an exchange between inclination and eccentricity for orbits above a critical inclination.
Orbits initially with high inclination can thus develop very high eccentricities.
For the case of a stationary axisymmetric quadrupole potential, the critical inclination is estimated to
be $ 39^{\circ}$ while  \cite{Ter2010}  and \cite{Tey2013} find  values as small as  $\sim` 20^{\circ}$
for a planet interacting with an axisymmetric disc in two and three dimensional treatments respectively.
In our case the disc is neither strictly axisymmetric nor steady. However, even though we find that it becomes
warped and unsteady from the viewpoint of the planet, the effect remains relatively small so that to a first approximation, the planet may be considered as moving in an axisymmetric potential. However, the component of angular momentum
parallel to the symmetry axis will not be strictly conserved,  as in that case, giving greater scope for generating inclination changes.

\begin{figure}
\centering
\includegraphics[width=5cm]{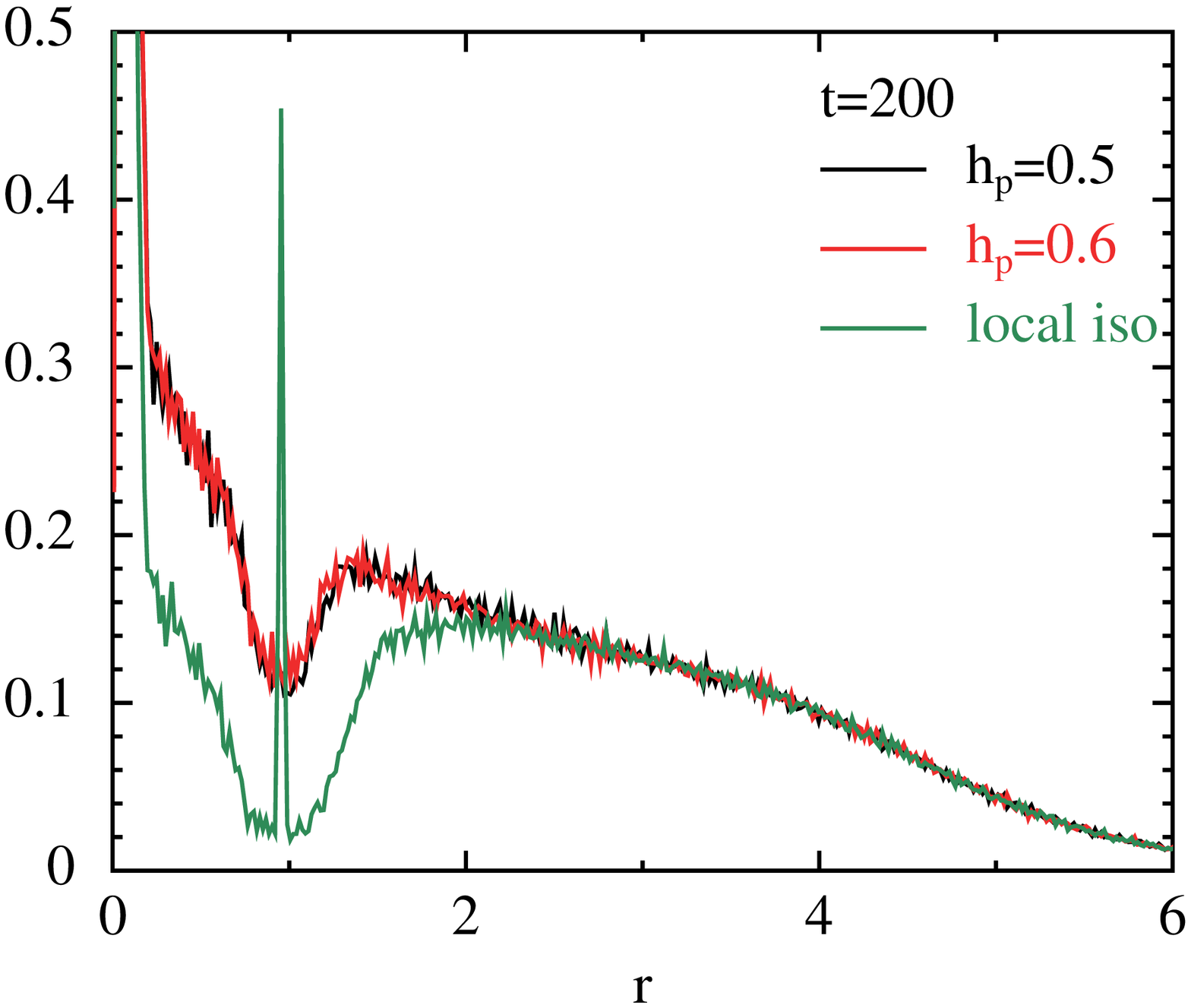}
\includegraphics[width=6cm]{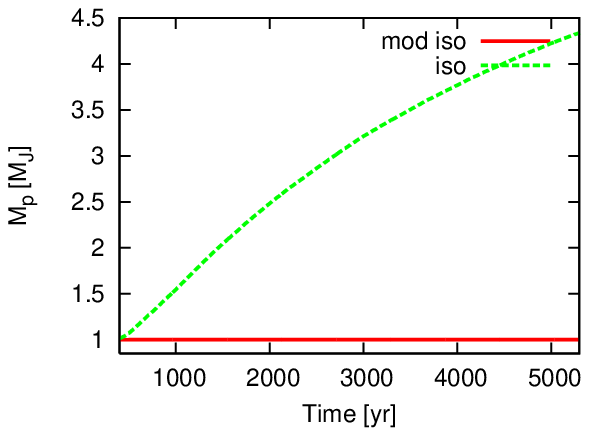}
\caption{ Upper panel:  gap formation for  $M_{p}=1~ \rmn{M_J}$ initially  with
different equations of state, the azimuthally averaged surface
density is shown after $200$ orbits. The two  upper curves corresponding to $h_p=0.5$ and $h_p=0.6$
are indistinguishable and the
lowest curve is for the locally isothermal equation of state.
Lower panel: mass growth
starting from an  initial planet mass $M_{p}=1\ \rmn{M_J}$ for $h_p=0.6$ and locally isothermal equation of state. The upper curve corresponds to the locally isothermal equation of state.
There is negligible accretion in the other case. }
\label{fig:gap_coplanar_EOS}
\end{figure}

 \begin{figure}
\centering
\includegraphics[width=6cm]{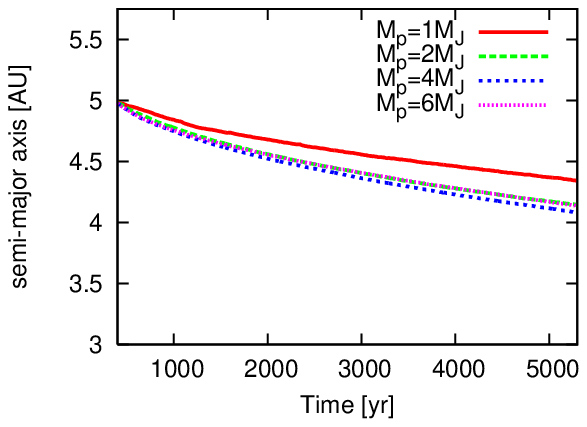}
\includegraphics[width=6cm]{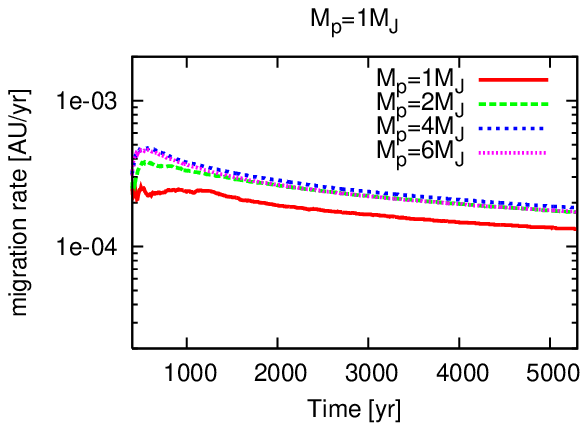}
\caption{Upper panel: Semi-major axis as a function of time  for coplanar planets of different masses (1, 2, 4, 6 $\rmn{M_J}$). Lower panel: cumulative mean  migration rates as a function of time }
\label{fig:a_coplanar}
\end{figure}

It is important to note that in the axisymmetric case, in the secular theory, 
a circular orbit at high inclination is on a separatrix \citep[e.g.][]{Fun2011}. Thus it becomes an unstable equilibrium point
and chaotic motion is to be expected in its neighbourhood.
 This raises a potential difficulty in identifying the cause of inclination changes
associated with high inclination near circular  orbits  as being entirely due to
frictional effects in simulation runs of limited length. However, we do not see significant eccentricity development, as
would be expected if a Lidov-Kozai effect operated in such cases,  which indicates that frictional effects are the main  cause of the evolution. 
In addition the characteristic times  for evolution can be compared to those 
expected from dynamical friction and they are found to be comparable.

\begin{figure*}
\centering
\includegraphics[width=6cm]{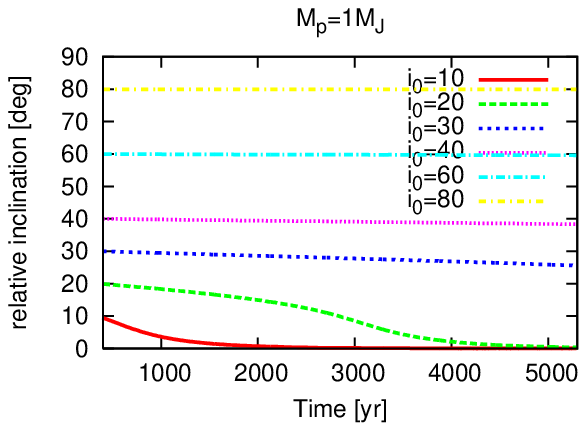}
\includegraphics[width=6cm]{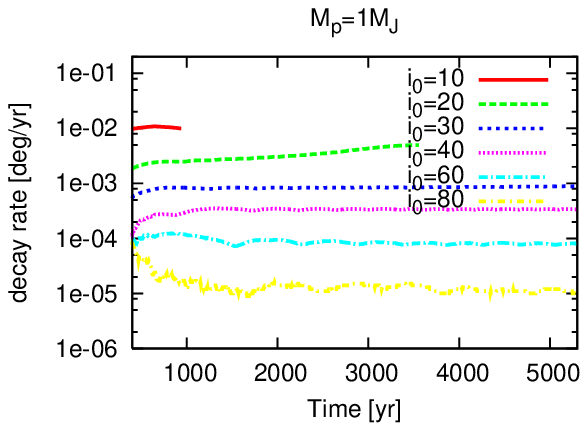}
\includegraphics[width=6cm]{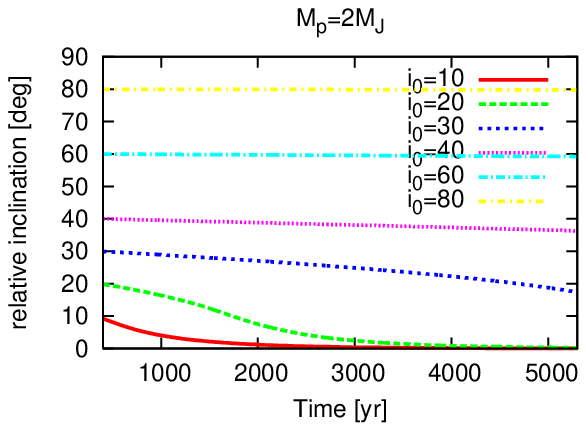}
\includegraphics[width=6cm]{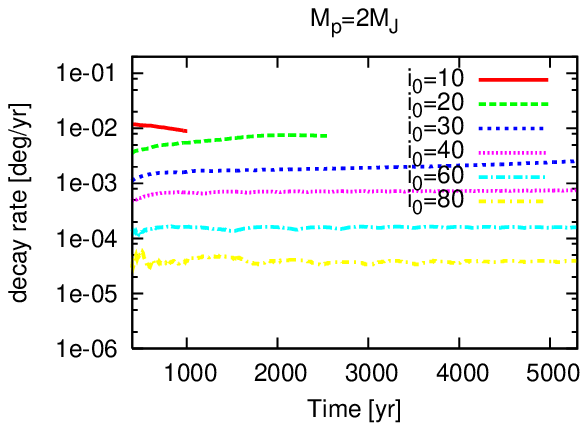}
\includegraphics[width=6cm]{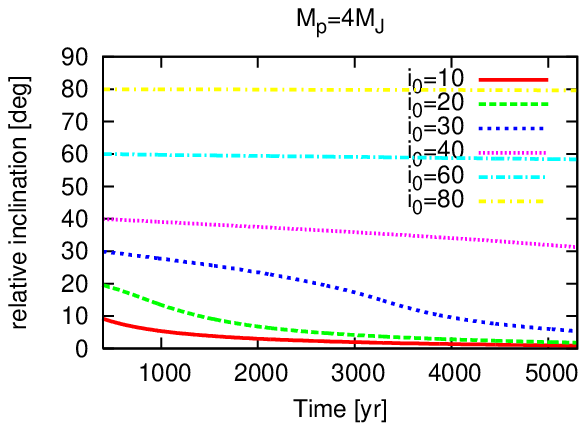}
\includegraphics[width=6cm]{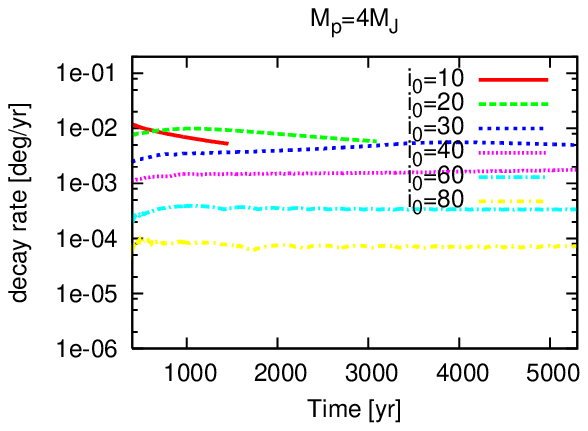}
\includegraphics[width=6cm]{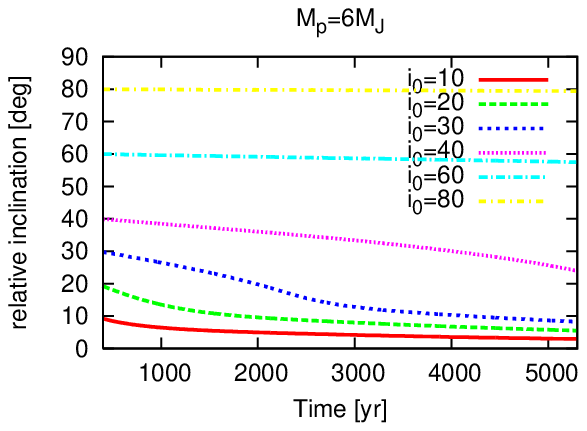}
\includegraphics[width=6cm]{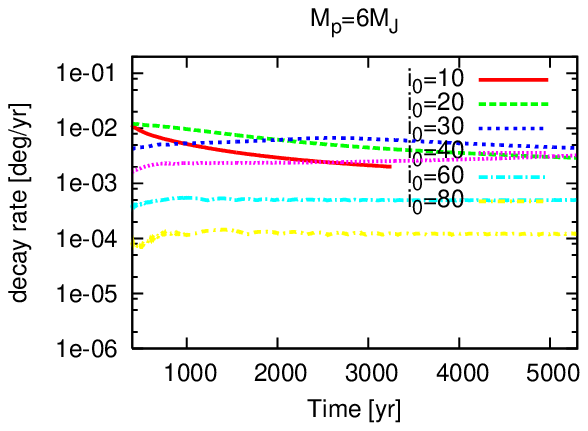}
\caption{Left: Relative inclination  for  planet masses (1, 2, 4, 6 $\rmn{M_J}$) as a function of time.
These were initiated on prograde circular orbits with initial relative inclinations
$10^{\circ}, 20^{\circ}, 30^{\circ},  40^{\circ}, 60^{\circ} $ and $80^{\circ}$.
  Right: Corresponding
 cumulative means of the  relative inclination decay rates as a function of time.}
\label{fig:rel_i}
\end{figure*}

\section{Numerical results with a coplanar planet} \label{sec:coplanar}
In order to understand the orbital evolution of massive planets in the presence of a disc
with its total angular momentum vector misaligned with that of the orbit,
it is important to consider the extent of the  occurrence of gap formation as this would expell material
from the neighbourhood of the planets location and so is expected to affect the strength
of the interaction of the planet with the disc. This is implied by the simple dynamical friction
estimate  for  the force exerted between the planet and disc given by (\ref{dynf}), which indicates that
this is proportional to the local disc  
density. Gap formation is most effective when the planet orbit and disc are coplanar as then the magnitude of the relative velocity between the planet and disc material in its neighbourhood is the smallest.
In addition, in contrast to the case of high orbital inclination, 
the planet then spends most of its time close to disc material. Accordingly, we begin by studying the
coplanar case.

\subsection{Gap formation in the coplanar case}\label{5.1}
In this section, we consider
the orbital evolution of  planets  with masses of  1, 2, 4 and 6 $\rmn{M_J}$
initiated in coplanar circular orbits with an initial semi-major axis of $5$ AU.
Fig. \ref{fig:gap_coplanar} shows the azimuthally averaged  disc surface density profile
after 200 planet orbits with the modified
equation of state. The development of a gap with a depth that increases with planet mass
is clearly apparent. The viscous criterion for gap formation given  by
\cite{Lin1993} is that  $M_p/M_* > 40\alpha_{SS}h_s^2$, where for convenience, we have assumed that the gap extends beyond the region where the modified equation of state changes it from being locally isothermal
and that dissipative processes are modelled by a \cite{Sha1973} $\alpha_{SS}$ parameterization. For the simulations presented here, this has been estimated to be in the range
$0.02-0.03$ (see appendix \ref{ap:ring_test}).  Thus we should require $M_p/M_*  > 0.002- 0.003.$
This is consistent with Fig.  \ref{fig:gap_coplanar}, which indicates a gap depression  of up to  $66\%$
for $M_p/M_* =  0.002$ and a  deep  gap for $M_p/M_* >  0.004$. The $1 \rmn{M_J}$ case is associated with a
partial gap associated  with a $33\%$ depression of the azimuthally averaged  surface density.
In addition, we remark that \cite{Cri2006} have studied the gap profile for different viscosities
and planet masses.
For $M_p=1\ \rmn{M_J}$, they find a gap  depth  of $\sim 40\%$  for $h_s =0.05$ and $\alpha_{SS}=0.025$.
  We remark that  the  whole  gap width we obtain, in this case  being about $0.6$ of the orbital radius,
is very similar to that found by \cite{Cri2006}.
 Our results are also consistent with those of \cite{Val2006} that indicated a somewhat less depleted gap found  in SPH simulations as compared to grid based simulations. But note that the treatment of the equation of state and accretion of gas particles differs from ours.

In order to illustrate the effects of the choice of the equation of state on the gap formation process and also
the accretion of material by the planet, we illustrate a comparison between  runs undertaken with  the locally isothermal equation of state and the  modified equation of state with $h_p=0.5$ and  $h_p=0.6$  in Figure \ref{fig:gap_coplanar_EOS}. 

In the locally isothermal case, the planet mass increases to over $4\ \rmn{M_J}$ on a time scale of about  five hundred orbits. 
In this case,  a high density of 
gas particles is able to accumulate in the neighbourhood of  the massive planet due to the small effective sound speed around the planet
resulting in relatively small pressure forces. 
Inside the Hill radius, SPH particles build up a hydrostatic equilibrium state with 
 a  very high mass density  peak centred on  the planet. 
The outer accretion radius is much smaller than  the Hill radius 
in order to prevent any interference with the dynamics on that scale. 
Therefore, only particles that come very close to the planet can satisfy the conditions to be accreted.

In contrast to the locally isothermal equation of state, the modified  equation of state
 prevents gas particles from accumulating in  the vicinity of the  planet due to the
significantly increased pressure forces  occuring because of the increased  sound speed.
 As a consequence, the Hill region around the planet does not contain a high density peak  
and  accretion onto the planet is less significant.

The results for $h_p=0.5$ and  $h_p=0.6$ are almost identical for the duration of the runs.  
Values $h_p < 0.5$ were  not used for $1\ \rmn{M_J}$ since they result in regions around the planet where the modified $c_{s,mod}$ is smaller
 than the $c_{s,iso}$ as discussed in appendix \ref{ap:peplinski}.
It can be inferred from this study that the simulations shown in this paper that adopt the modified   equation of state  do not require implementation of the accretion
algorithm.
 It is nonetheless   retained  for completeness and in order to ensure that any  potential singularities at the locations of  massive particles can be dealt with.

\subsection{Migration in the coplanar case}
In Fig. \ref{fig:a_coplanar}, the evolution of the semi-major axis $a$ and the migration rate of a coplanar planet are studied.
In order to   eliminate any effects due to  short timescale  variations in defining migration rates,
we evaluate a cumulative mean  of $da/dt$ as
\begin{equation}
\frac{{\overline {d a}}}{dt} = \frac{1}{t}\int^t_0 - \frac{d a(t') }{dt'} dt'= \frac{a(0)-a(t)}{t}\ ,
\end{equation}
where  $a(t)$ is the semi-major axis at time $t$ and $a(0)$ is its value at  $t=0$.
As expected, the migration is always found to be inwards. 
At the start of the simulations, no gap is present in the disc.
During the gap formation  phase, the migration rates are largest. 
At later times, the planets have   opened  gaps  and the migration  gradually slows down.
The behaviour of the three largest masses is very similar. 
For $M_p=1\ \rmn{M_J}$, the gap is not as deep as in the other
cases resulting in less of a decrease of the migration rate as compared to the other masses.
At large times and for gap forming planets that are not too massive compared to the disc mass,
inward migration is expected to occur on the viscous timescale of the disc \citep{Lin1986}. 
The steady state inflow velocity in a viscous disc is $1.5\alpha_{SS}h^2_sr_s\Omega_s$ \citep[e.g.][]{Pri1981}.
This leads to an expected migration rate of $3\pi \alpha_{SS} h^2_s/P_{orb},$ where $P_{orb}$
is the local orbital period. For $ \alpha_{SS}=0.025,$ this corresponds to $2.5\times 10^{-4}$ AU/yr at $5$ AU.
This is in good agreement with the rates shown at the latest time  in Fig. \ref{fig:a_coplanar} for the three larger masses.
For $M_p=1\ \rmn{M_J}$, the migration rate is   slightly slower at this  time.
This may be on account  of the differing initial evolution, causing it to be at a larger radius
 and the presence of more material in the gap region.

\begin{figure*}
\centering
\includegraphics[width=5cm]{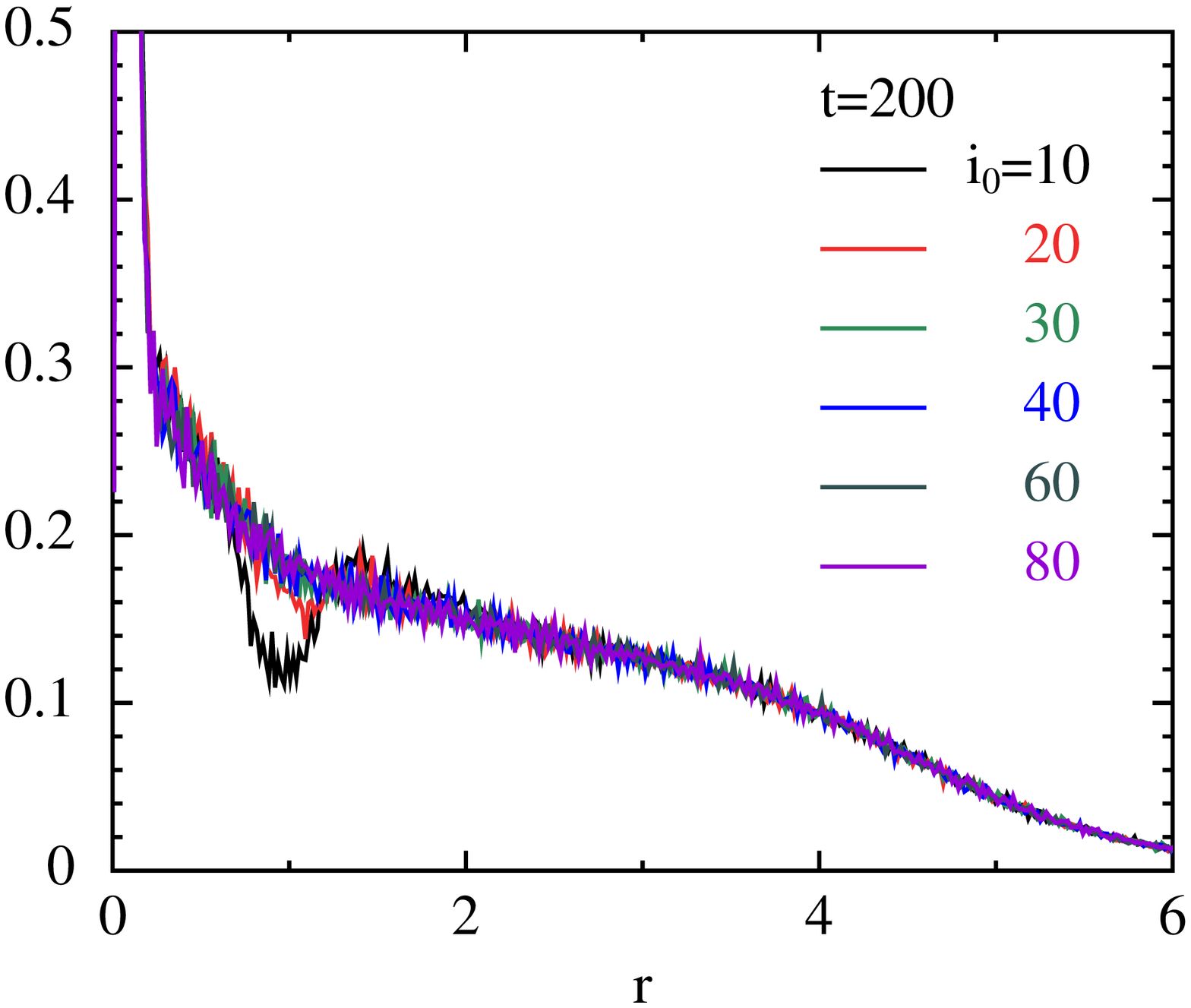}
\includegraphics[width=5cm]{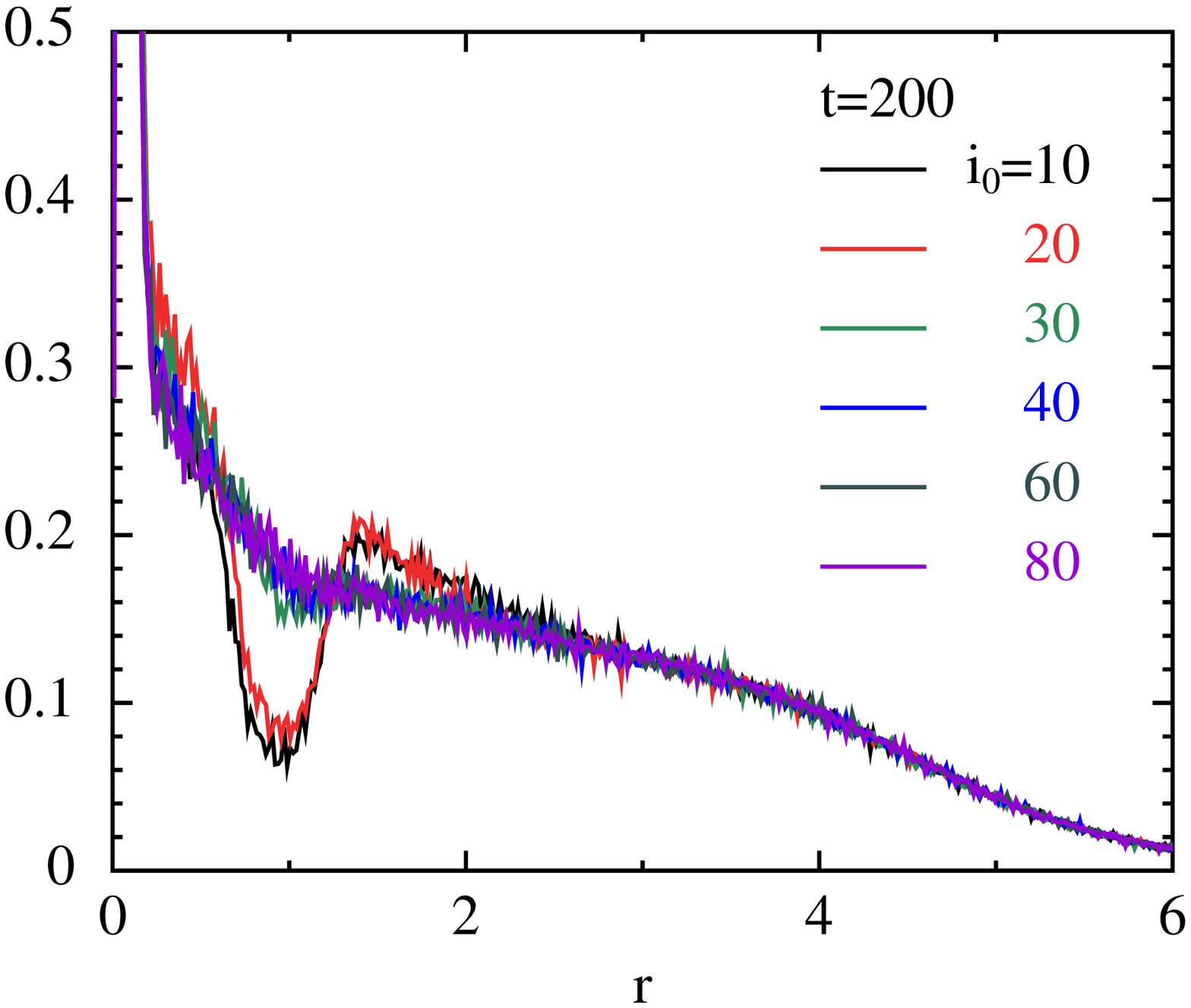}\\
\includegraphics[width=5cm]{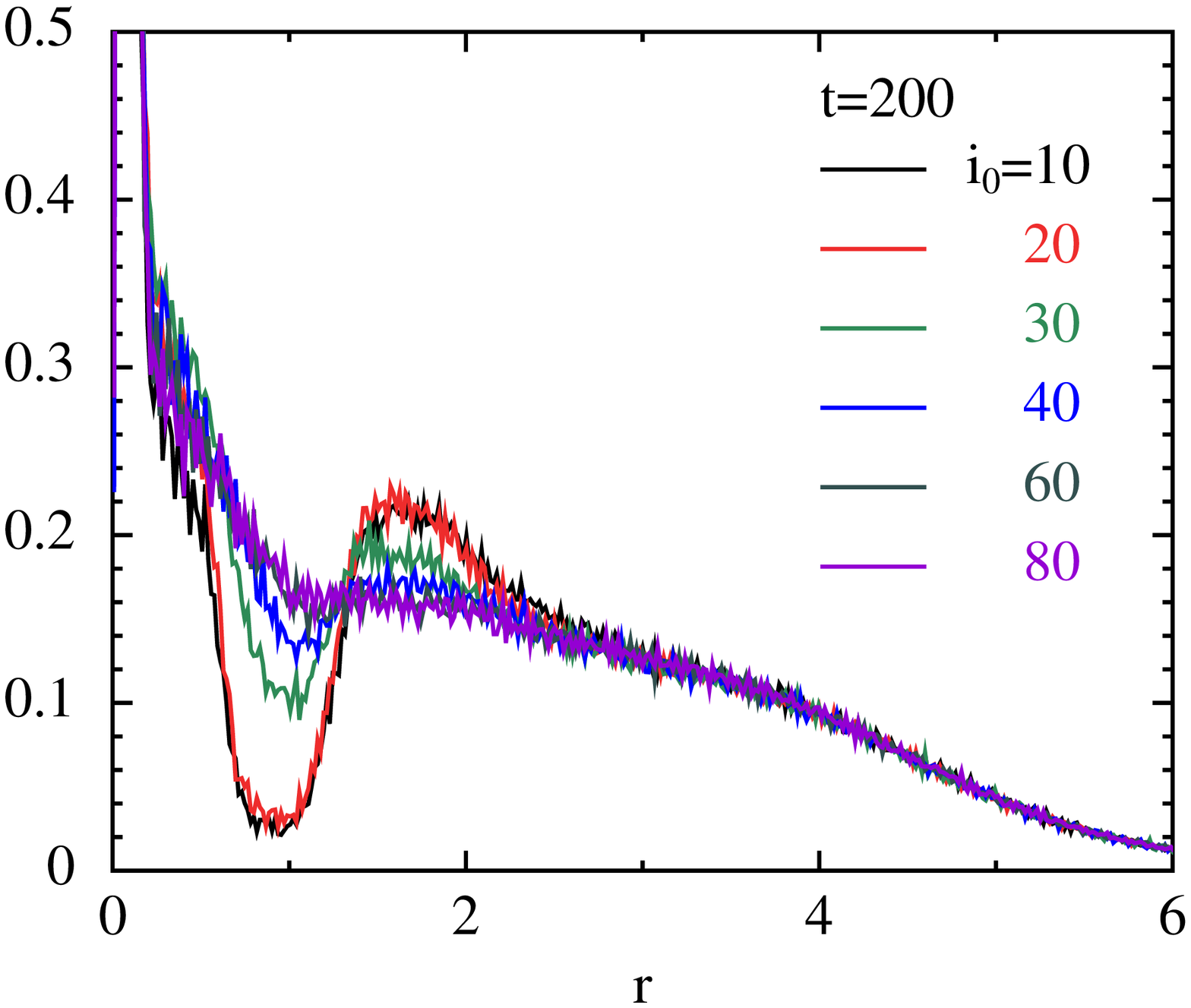}
\includegraphics[width=5cm]{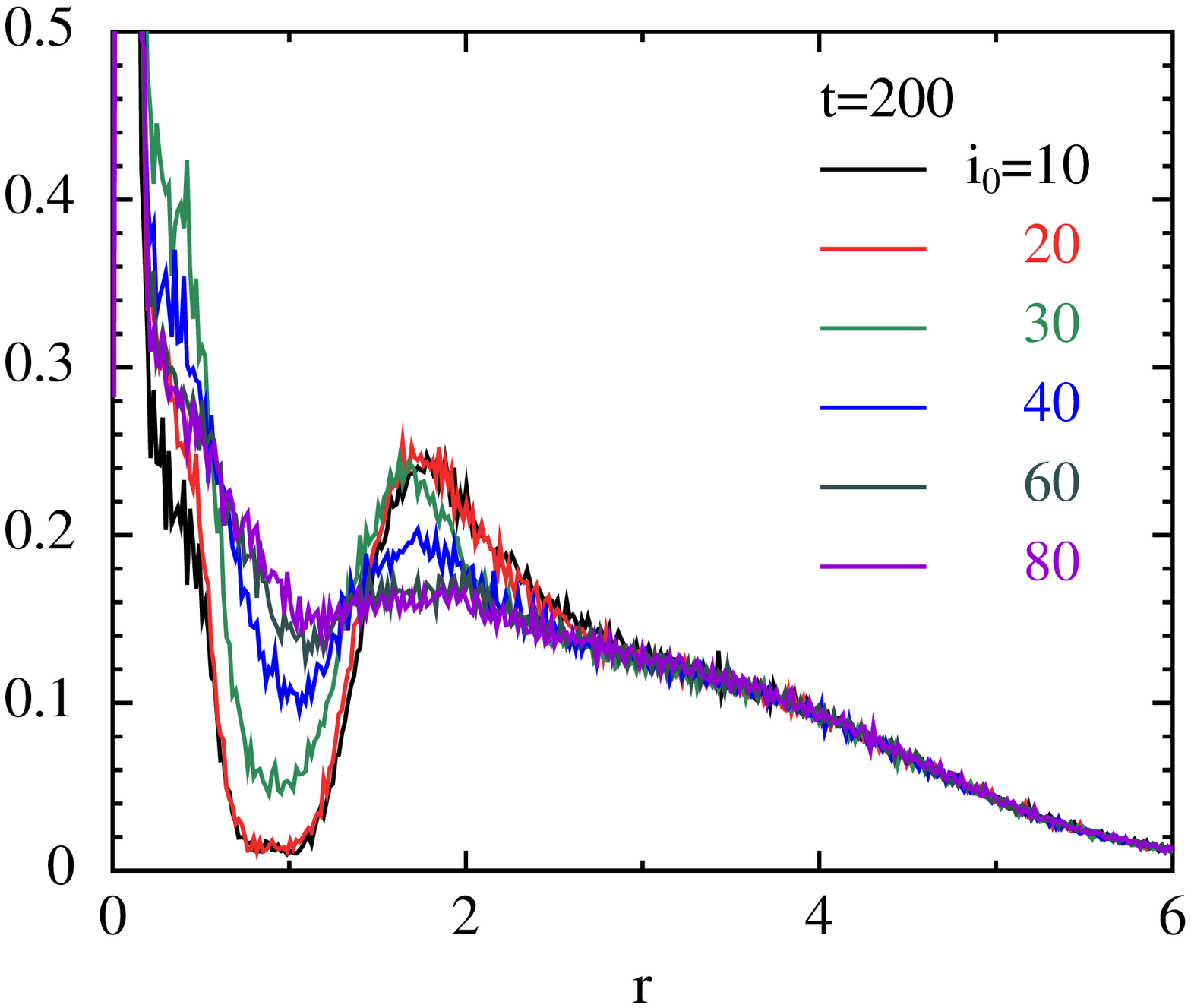}
\caption{Gap formation for different planet masses and initial relative orbital  inclinations:
The azimuthally averaged surface density is shown after 200 orbits
for, moving from left to right and top to bottom,
$M_p = 1, 2, 4$ and  $6\ \rmn{M_J},$  starting with different initial relative inclinations.}
\label{fig:gap_i}
\end{figure*}

\section{Interactions of planets with non zero inclination with the disc} \label{sec:inclined}
We now consider the evolution of planets  initiated on circular orbits with non zero relative inclination.
The relative inclination, $i$,  is defined through
\begin{eqnarray}
&i&= \arccos\left( {\frac{{\bf l}_{p} \cdot {\bf L}_{D}}{|{\bf l}_{p}| \cdot |{\bf L}_{D}|}}\right), \hspace{2mm} {\rm  with} \\
&{\bf L}_{D}&=\sum_i m_i ({\bf r}_i \times {\bf v}_i).\label{incdef} 
\end{eqnarray}
Here  ${\bf l}_{p}$ is the angular momentum vector of the planet and ${\bf L}_{D}$ 
is the total angular momentum vector of the disc,  with  the summation in (\ref{incdef}) being  taken over all active gas particles.
In simulations starting with circular orbits,  the planetary orbit develops only very small eccentricities over their run times.
This is in contrast to the situation that occurs  when orbits with high enough 
relative inclination  are initiated with a modest non zero
 eccentricity (see Section \ref{sec:incecc}).

\subsection{Decay of the inclination for prograde initially circular orbits}

Fig. \ref{fig:rel_i} shows the inclination $i$ of the planet orbit relative to the  plane normal to the
disc total angular momentum vector and the cumulative mean of its decay rate
for  planets of  mass $1, 2, 4$ and $6\ \rmn{M_J}$ 
initiated on prograde circular orbits with initial relative inclinations
$10^{\circ}, 20^{\circ} , 30^{\circ},  40^{\circ}, 60^{\circ} $ and $80^{\circ} .$ 
The cumulative mean of $di/dt$ was adopted in order to iron out small time scale variations. This was defined through 
\begin{equation}
{\overline \frac{d i}{dt}} = \frac{1}{t} \int^t_0 - \frac{d i(t')}{dt'}dt'=  \frac{i(0)-i(t)}{t}\ ,
\end{equation}
with $i(t)$  being the relative inclination at time $t$ and $i(0)$ its  initial value.
This approach loses meaning when the  planet becomes almost coplanar with the disc.
As can be seen from  the left hand panels of Fig. \ref{fig:rel_i}, the latter  readily happens for $i_0 <  20^\circ.$ 
Cumulative means , defined as above would  then approach $i(0)/t,$ which is determined by the initial condition.
Accordingly  decay rates are only shown for relative inclinations above $5^\circ$ so avoiding this limit. 

\begin{figure*}
\centering
\includegraphics[width=6cm]{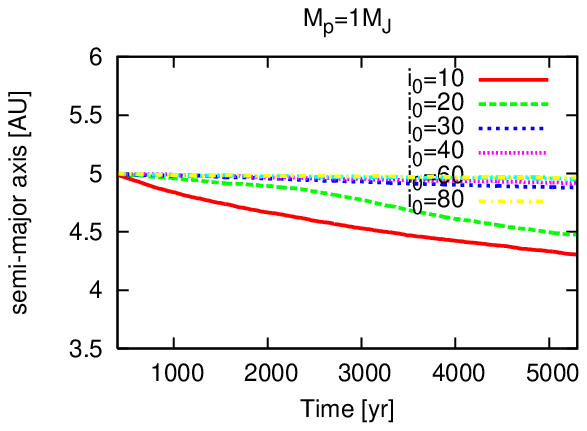}
\includegraphics[width=6cm]{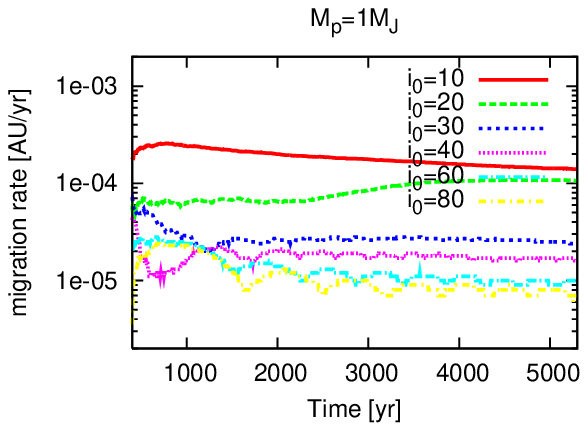}
\includegraphics[width=6cm]{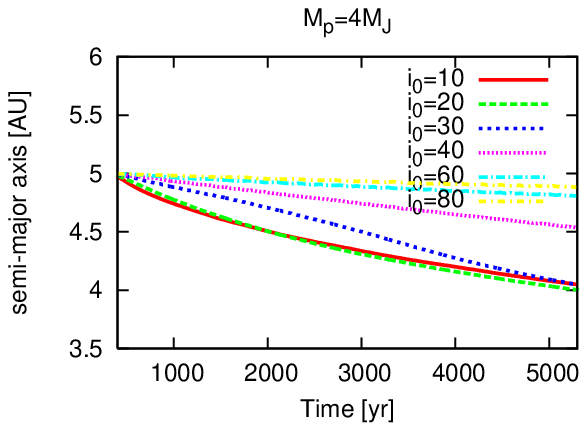}
\includegraphics[width=6cm]{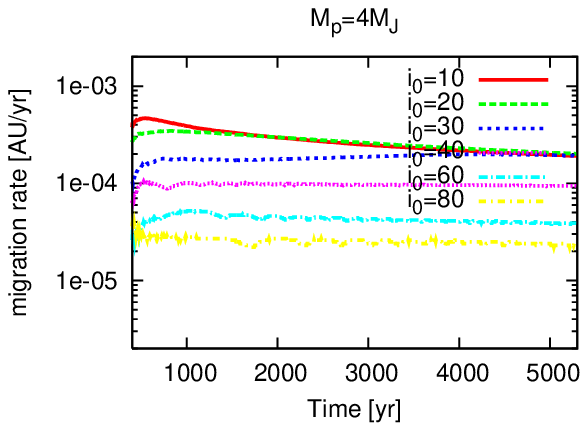}
\caption{Left: Evolution of the semi-major axis $a$ for  planets  with $M_p = 1$ and $4  \ \rmn{M_J}$
for a range of  initial relative orbital inclinations, Right: cumulative mean decay rates.}
\label{fig:a}
\end{figure*}
 
In general, the cumulative mean of the inclination decay rate decreases in magnitude as the inclination increases.
This  qualitative feature is consistent with the  simplistic 
estimate of the orbital evolution time scale based on dynamical friction
considerations  given by (\ref{TD}).
This  equation suggests that this decay rate should be proportional to $i/(\sin i \sin (i/2)^{3}).$
Thus comparing the early cumulative mean of the  decay rates for  $i= 80^{\circ}$ and $i= 10^{\circ}$ for  $1\ \rmn{M_J}$
in Fig. \ref{fig:rel_i}, we find a ratio of $\sim 10^{-3},$ with (\ref{TD}) indicating  a ratio of $\sim 3\times 10^{-3}.$  
 This ratio increases for larger masses on account
of gap formation at low inclination. However, in all cases the ratio of the decay rates for  $i= 40^{\circ}$ and $i= 80^{\circ}$
where gap formation is less relevant is  $\sim 16.$ In this case, (\ref{TD}) indicates a value of $\sim 5.$
In view of the  very rough nature of the estimate given by (\ref{TD}) the level of agreement is satisfactory.

For $i_0=10^\circ$ the results given in Fig. \ref{fig:rel_i} indicate that  $T_D \sim 10^3$~yr in the case of   $1\ \rmn{M_J}$
while for $i_0=80^\circ$,  $T_D \sim 10^7$~yr initially.  But note that the
decay rate accelerates considerably as the inclination decreases so that for $i_0=40^\circ$,  $T_D \sim 2\times 10^5$~yr .
Again this is in reasonable agreement with the estimate given above derived from equation (\ref{TD}).

For small to intermediate $i$ and larger planet masses,  the $i$-decay is strongly dependent on  gap formation,
$i$ decreases at an accelerating rate  towards smaller values as long as no gap is formed.  As soon as $i$ is sufficiently small to open a gap in the disc,  the rate of inclination decrease  slows down.  At large inclinations, the inclination decay rate increases
with planet mass  so that for $i_0=80^\circ$, the decay rate is an order of magnitude faster for $6\ \rmn{M_J}$  as compared to $1\ \rmn{M_J}$.

\subsection{The dependence of gap formation on  orbital inclination for prograde orbits} 
Because the relative velocity between the planet and disc material increases with orbital inclination,
the interaction becomes weaker and gap formation is less likely.
There can be a threshold  inclination above which gap formation does not occur.
This threshold inclination for gap opening increases with planet mass.
Fig. \ref{fig:gap_i} shows the azimuthally averaged disc surface mass density after 200 orbits for different planet masses and initial inclinations $i_0.$
We remark  that,  as can be deduced from the results presented  above,
 after 200 orbits the relative inclination $i$  will have  decreased significantly from its initial value $i_0$ in some cases. 
For very high inclinations, $\Sigma$ is marginally  perturbed for $M_p= 6\ \rmn{M_J}$, 
the perturbation being  weaker for the smaller masses as expected. 
With decreasing $i_0$, the interaction between the planet and the disc becomes stronger  
until eventually the  threshold below which a noticeable gap  starts to form is passed.  
Thus 
for $M_p = 1\ \rmn{M_J}$, only the $i_0=10^\circ$ curve is able to open a noticeable partial gap.
For $M_p= 2\ \rmn{M_J}, 4\ \rmn{M_J}$  and $6\ \rmn{M_J},$ the threshold initial inclinations are  $i_0=20^\circ, 30^\circ$ and $40^\circ$, respectively.

\begin{figure}
\centering
\includegraphics[width=6cm]{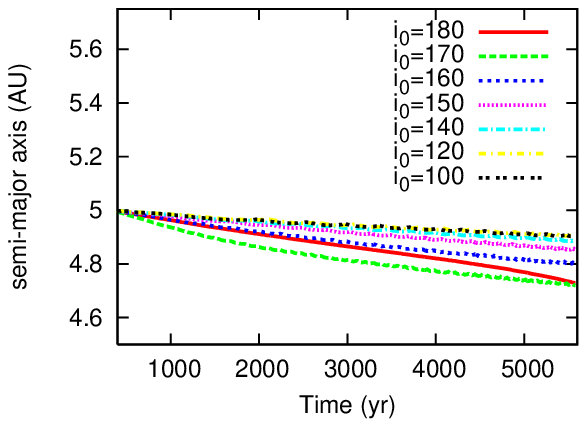}
\includegraphics[width=6cm]{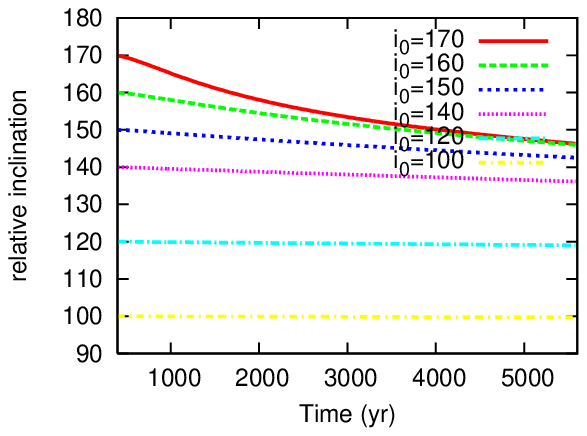}
\caption{Orbital evolution  of
a planet with $M_p=4\ \rmn{M_J}$ starting on retrograde circular orbits with different
initial relative inclinations. In the upper panel, the semi-major axis is shown as function of time.  
Lower panel: relative inclination as function of time.}
\label{fig:retrograde}
\end{figure}

\subsection{Evolution of the semi-major axis for finite relative orbital inclination}
Fig. \ref{fig:a} shows the evolution of the semi-major axes and 
cumulative means of the migration rates for planets with $M_p = 1$ and $ 4\ \rmn{M_J}$
initiated on circular orbits with different initial relative orbital inclinations. The cases $M_p=2$ and $6\ \rmn{M_J}$ show very similar behaviour, therefore will not be shown.
The runs shown  are identical to those illustrated in Fig. \ref{fig:rel_i}. Thus
corresponding relative inclinations can be read off.
As in the coplanar case,  the migration is always inwards and the orbits remain
almost circular for the run times shown. 
For planets starting on inclined orbits, the migration rate is slower than for the coplanar case
 due to the reduced interaction of the planet with the disc.
When the relative orbital inclination approaches zero, it is found that
 the migration rate approaches the value appropriate to coplanar planets.
For example, for $M_p= 1\ \rmn{M_J}$ and $i_0=20^\circ$,  $i$ decreases to  
$<5^\circ$ after $\sim$ 3000 years.
During this  time interval, the migration rate approaches the value for a planet of the same mass with  $i_0=0^\circ.$
For  $M_p = 4\ \rmn{M_J}$ and $i_0=30^\circ$, the relative inclination decreases to  below $10^\circ$ for  $t>3000\ \rmn{yr}.$ 
At this stage the migration rate approaches that for the coplanar case.

\begin{figure}
\centering
\includegraphics[width=7cm]{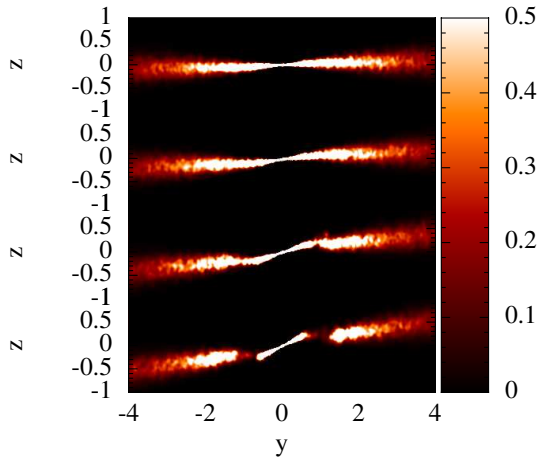}
\caption{Cross section of the disc as seen in the plane  $x=0$  after 200 orbits,  for
 $M_p = 1, 2, 4$ and  $6\ \rmn{M_J},$ and $i_0=30^\circ$. $y$ and $z$ in units of  $5\ \rmn{AU}$. The orientation of the coordinate system is such that the $z$ axis corresponds to the direction of the angular momentum vector of the initial disc and the planet is initiated at $y=0$ with $x>0$ and $z>0$. [All SPH cross section plots were created using SPLASH \citep[][]{Pri2007}] 
}
\label{fig:warp_30_0}
\end{figure}
\begin{figure}
\centering
\includegraphics[width=7cm]{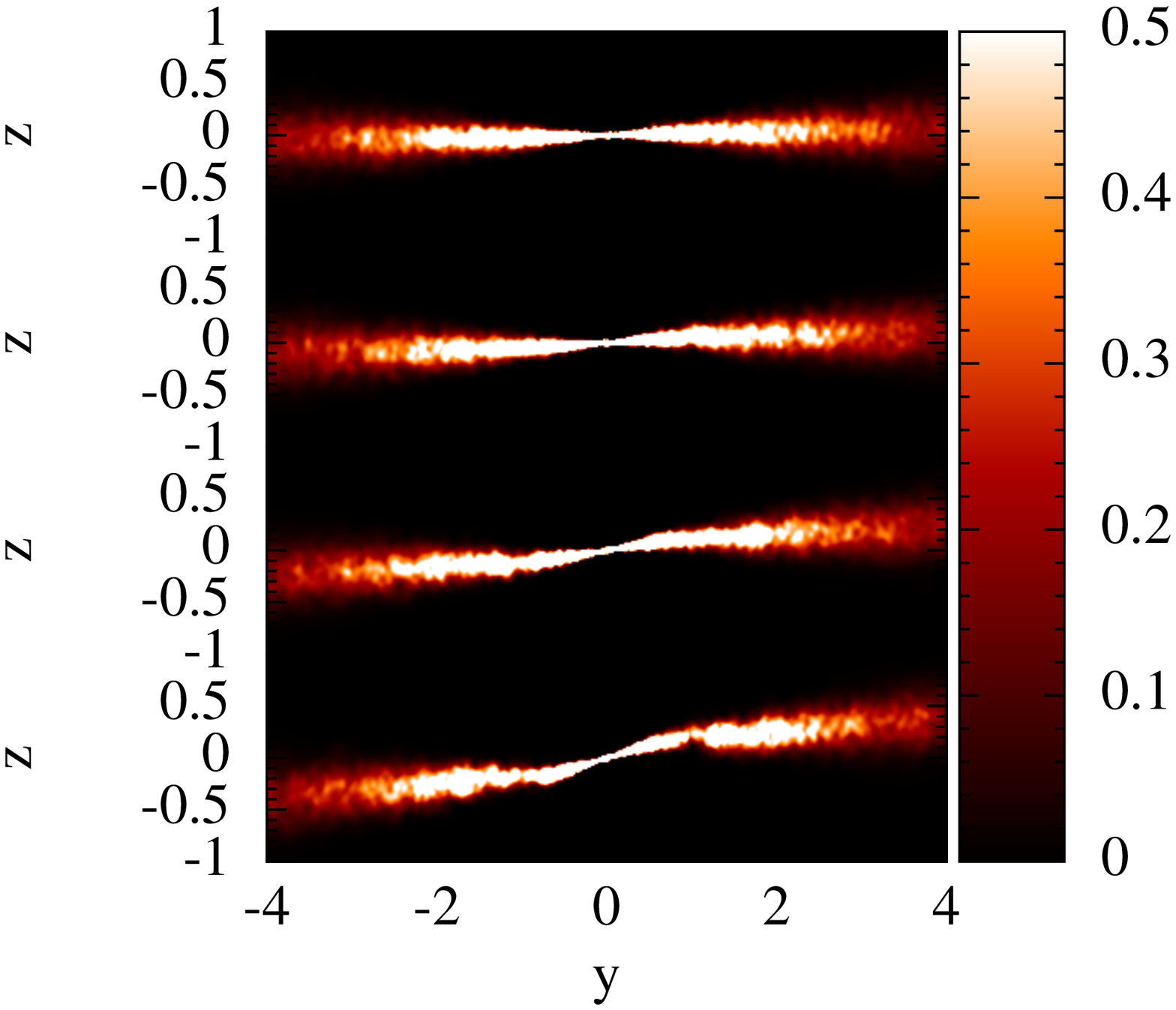}
\caption{Cross section of the disc as seen in the plane  $x=0$  after 200 orbits,  for
 $M_p = 1, 2, 4$ and  $6\ \rmn{M_J},$ and $i_0=40^\circ$. The coordinate system is as defined in Fig. \ref{fig:warp_30_0}. }
\label{fig:warp_40_0}
\end{figure}
\begin{figure}
\centering
\includegraphics[width=7cm]{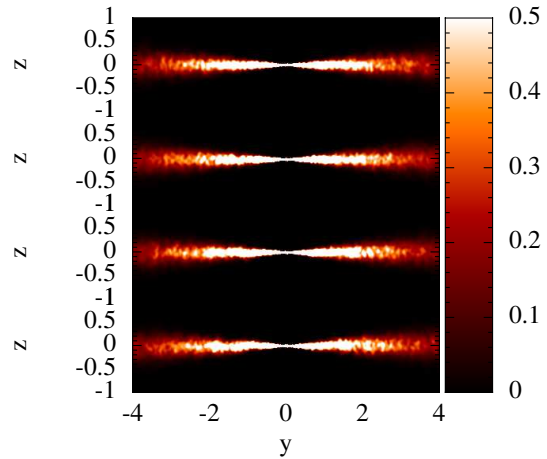}
\caption{Cross section of the disc as seen in the plane  $x=0$  after 200 orbits,  for
 $M_p = 1, 2, 4$ and  $6\ \rmn{M_J},$ and $i_0=80^\circ$. The coordinate system is as defined in Fig. \ref{fig:warp_30_0}.}
\label{fig:warp_80_0}
\end{figure}

\subsection{High inclination retrograde orbits}
 Finally in this section, we give  examples of the orbital evolution
of a planet starting on a retrograde circular orbit  with varying initial  relative  
orbital inclination. In Fig. \ref{fig:retrograde}, we illustrate the evolution of the
semi-major axis and the relative orbital inclination as functions of time for a planet of mass $M_p=4\ \rmn{M_J}.$
We remark that the direction of evolution of the inclination is always towards coplanarity $(i_{rel}=0^\circ),$
even when  the orbit starts out with  $i_0 \sim 180^{\circ}.$ This is to be expected from a frictional
interaction between the planet and the disc which tends to communicate angular momentum in the direction of the disc's angular momentum vector to the orbit of the planet.
In all of these cases, the orbit remains approximately circular.
It is seen that the most rapid evolution occurs for the largest inclinations for which
the planet tends to become embedded in the disc and thus undergo a more sustained frictional interaction.

\begin{figure}
\centering
\includegraphics[width=6cm]{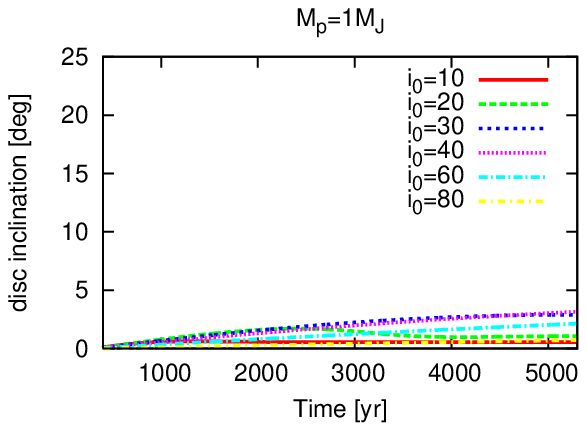}
\includegraphics[width=6cm]{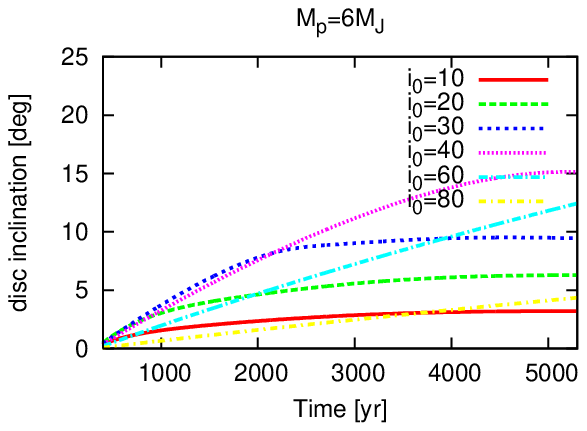}
\caption{Evolution of total disc inclination for the first $5000\ \rmn{yr}$ for different initial inclinations in the cases $M_p=1$ and $6\ \rmn{M_J}$.}
\label{fig:disc_inclination}
\end{figure}
\begin{figure}
\centering
\includegraphics[width=7cm]{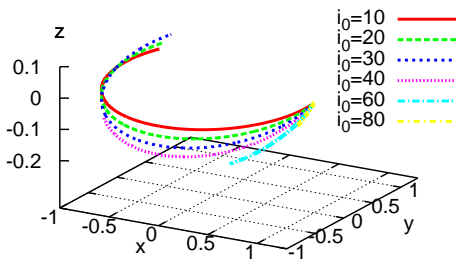}
\caption{Time-dependent evolution of the unit vector ${\bf L}_{D,tot}$ for  $M_p=4\ \rmn{M_J}$ and different initial
relative  inclinations for a period of  $\sim 5000$ years. The coordinate system is as defined in Fig. \ref{fig:warp_30_0}.} 
\label{fig:disc_precession}
\end{figure}

\section{The response of the disc: inclination changes, warping and precession} \label{sec:response}

As a response to a massive planet, a circumstellar disc can change its orientation significantly,
developing a warped structure.
So far, we have considered the inclination of the planetary orbits 
with respect to the plane
for which the angular momentum vector of the disc as a whole is parallel to the normal.
However,  warping can be present in the disc, dividing it into two parts with angular momentum vectors 
pointing in different directions and thus associated with different inclinations.
 In Figure \ref{fig:warp_30_0}, we show a cross section consisting  of the density distribution 
as seen in the plane $x=0$ 
for different planet masses, all of which started with an 
initial inclination, $i_0=30^\circ,$ after 200 orbits.
This value of $i_0$ was chosen  as maximum warping is expected for an inclination intermediate
between $0^{\circ}$ and $90^{\circ},$ there being none at these extremes. 
It is seen that a visible warped inner section of the disc  is produced by the more massive planets while planets 
with $M_p=1$ and $2\ \rmn{M_J}$ do not succeed in creating a  visibly warped disc. 
The same result is found for an initial inclination $i_0=40^\circ$ 
as is illustrated in Fig. \ref{fig:warp_40_0}.
The disc profile is shown for  $i_0=80^\circ$  for the
different planet masses in Fig. \ref{fig:warp_80_0}. 
 In this case no warp is produced since the planets orbit is almost perpendicular to the disc mid plane. 
An indication of the degree of warping in the above cases can be obtained by measuring the
 inclination angles   between the inner and outer discs as seen in these cross sections.
It is noticeable  that the inclination of the inner disc can be quite substantial
in the direction of aligning with the inclination of the planetary orbit
and that this  effect is much less severe for the outer disc.
Table \ref{tab:warp_angles}  lists quantities characterising the disc warp after 200 orbits  for three runs.
\begin{table}
\centering
\begin{tabular}{|c|c|c|c|c|}
\hline
$M_p$ $[\rmn{M_J}]$ & $i_0$ [deg] & $i_{rel}$ [deg] & $\alpha_{warp}$ [deg] \\
\hline
4 & 30 & 17.0 & 10.4 \\
\hline
6 & 30 & 14.6 & 19.1 \\
\hline
6 & 40 & 34.8 & 10.7 \\
\hline  
\end{tabular}
\caption{Characterising the disc warp: 
the first column gives the planet mass, the second column the initial relative orbital inclination,
the third gives the  relative orbital inclination after 200 orbits
and the fourth column gives the difference in the inclination angles
of the inner and outer disc to the initial disc mid plane.}
\label{tab:warp_angles}
\end{table}
As most of the inertia of the disc is in its outer parts,
they dominate the inclination of the disc defined by considering its
total angular momentum.
Fig.  \ref{fig:disc_inclination} shows the evolution
of the inclination of the disc,  defined in this way, 
as a function of initial relative  inclination of its orbit for the two extreme cases $M_p=1$ and $M_p=6\ \rmn{M_J}$. 
By definition  the disc inclination is zero at the start of the simulations.
The total disc inclination is determined  from
\begin{eqnarray}
i_{D}=\arccos \left( \frac{L_{D,z}}{|{\bf L}_ {D}|} \right)\ .
\end{eqnarray}
In general, the disc inclination based on its total angular momentum does not attain large values in any case because of the high disc inertia.
For all the planet masses, the effect of the planet is strongest for the intermediate value  $i_0=40^\circ$ 
as expected. Simulations with $i_0= 60^\circ$ and $i_0= 80^\circ$
produce lower disc inclinations as would be expected for a rigid planar disc because
there would be reduced precessional torques in those cases.
For the highest mass  of $6\ \rmn{M_J}$, the total disc inclination only  attains values  up to $15^\circ$.

\begin{figure}
\centering
\includegraphics[width=6cm]{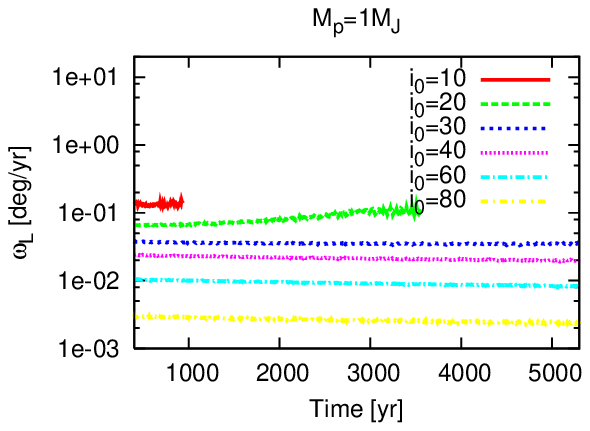} 
\includegraphics[width=6cm]{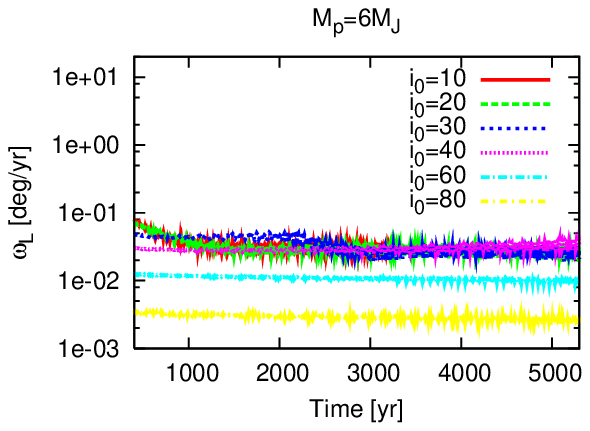}
\caption{The magnitude of the  retrograde precession  angular velocity of the disc 
found from equation (\ref{omegaL}) as a function of time for  planet masses  with $M_p = 1$ and $ 6\ \rmn{M_J}$ and various initial relative inclinations.}
\label{fig:disc_precession_velocity}
\end{figure}

\subsection{Disc precession}
The gravitational interaction of  a disc with a planet in an inclined orbit  is expected to make the orbit precess about the total angular momentum vector of the system.
Provided communication across the disc through either wave propagation or viscous diffusion occurs more rapidly than the precession rate, the response of the disc would be expected to be to precess  like a rigid body
in a retrograde sense relative to the orbit \citep{Lar1996} while simultaneously undergoing a non disruptive warping.
Due to angular momentum conservation of the composite system, this precession
is also expected to be about  the total  angular momentum vector.

In Figure \ref{fig:disc_precession}, we show the trajectory of the vector ${\bf L}_{D,tot}$ defined as
\begin{eqnarray}
{\bf L}_{D,tot}=\frac{{\bf L}_{D} \times {\bf L}_{tot}}{|{\bf L}_{D} \times {\bf L}_{tot}|} \ .
\end{eqnarray}
Thus ${\bf L}_{D,tot}$ is  a unit vector in the direction
normal to the plane containing  the disc angular momentum, ${\bf L}_D$, and
 the total angular momentum vector, ${\bf L}_{tot}={\bf l}_p + {\bf L}_D$, of the system.
We focus on  the case with $M_p=4\ \rmn{M_J},$ noting that  other masses  cases manifest similar behaviour. 
When there is rigid body precession, ${\bf L}_{D,tot}$ rotates uniformly with the precession
frequency about ${\bf L}_{tot}$ also in the retrograde sense.
The magnitude of the  angular velocity of ${\bf L}_{D,tot}$ can be found from
\begin{eqnarray}
\omega_L=\left| \frac{{\bf L}_{D,tot} \times {\bf v}_L}{|{\bf L}_{D,tot}|^2} \right| \ , \label{omegaL}
\end{eqnarray}
where ${\bf v}_L= dL_{D,tot}/dt.$
This is shown  as a function of time in Figure \ref{fig:disc_precession_velocity} for the  runs
with planet masses $M_p=1,$ $6\ \rmn{M_J}$ and initial relative orbital inclination.
It is found to be almost constant in all cases when a significant relative inclination is present
that does not decrease significantly  with time,
which is a good  indication of quasi rigid body precession.
It decreases in magnitude as the relative inclination
increases which is expected as a reflection of the decrease in magnitude of the precessional torque
acting between the planet and the disc as the relative inclination increases.

\subsection {Precession of the line of nodes of inclined  prograde planet  orbits}
In addition to the disc precession, the planet orbit also precesses.
The orbital angular momentum vector should precess 
about the total angular momentum vector with the same angular velocity. 
\begin{figure}
\centering
\includegraphics[width=7cm]{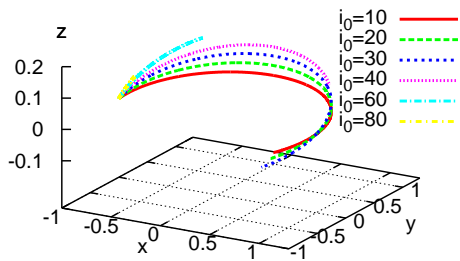}
\caption{Time-dependent evolution of the precession vector ${\bf P}(t)$ for $M_p=4\ \rmn{M_J}$ 
and different initial inclinations $i_0$ over $\sim 5000\ \rmn{yr}$. The coordinate system is as defined in Fig. \ref{fig:warp_30_0}.
}
\label{fig:precession}
\end{figure}
Figure \ref{fig:precession} shows the evolution of the precession vector defined as
\begin{eqnarray}
{\bf P}=\frac{{\bf l}_{p} \times {\bf L}_{tot}}{|{\bf l}_{p} \times {\bf L}_{tot}|}
\end{eqnarray}
for the case $M_p=4\ \rmn{M_J}$ and different initial inclinations for a time period of $\sim 5000\ \rmn{yr}$.
We remark that ${\bf P}$ is a  unit vector perpendicular to the plane containing
the orbital angular momentum vector
and the total angular momentum vector.
It is found to rotate with the same precession precession frequency as the disc,
as would be expected if the latter precessed like a rigid body .
The same behaviour has also been found for all other  planet mass cases studied.
The precession is retrograde  with  the  inclinations of the plane in which ${\bf P}$ evolves being dependent 
on the initial relative inclination of the planet.
 The higher the relative initial inclination of the planet orbit,  the higher  the inclination of the plane of ${\bf P}$. 
The angular velocity of orbital precession coincides exactly with that of the disc.


\section{Evolution of the eccentricity and the indication of a Lidov-Kozai effect at high inclination}\label{sec:incecc}
So far, we have described simulations for which planets were initiated on circular orbits
and in all cases the orbital eccentricity remained very small for the duration 
of the simulations.
\begin{figure}
\centering
\includegraphics[width=6cm]{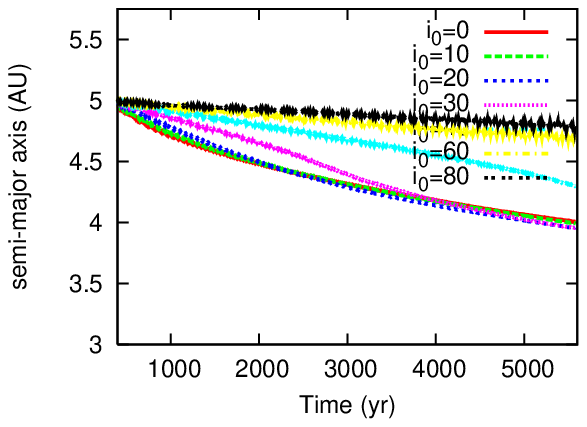}
\includegraphics[width=6cm]{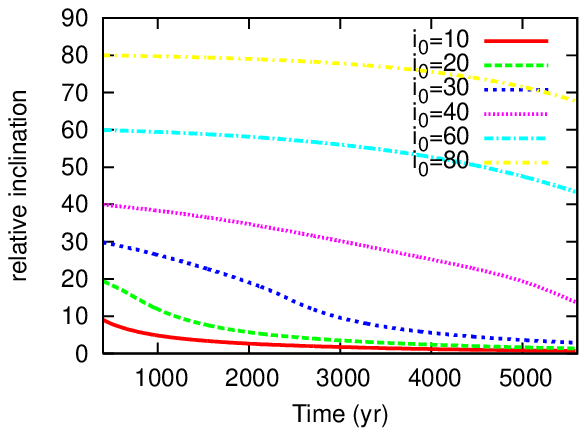}
\includegraphics[width=6cm]{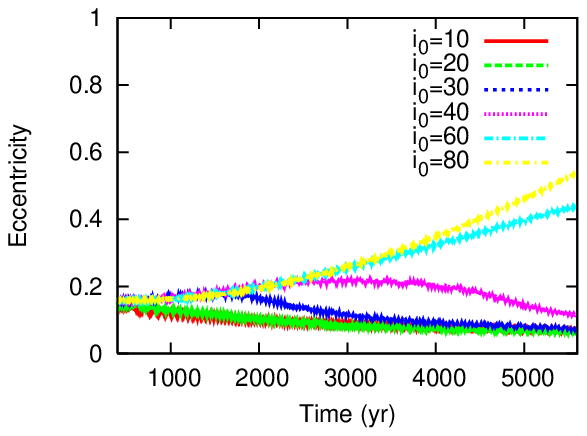}
\includegraphics[width=6cm]{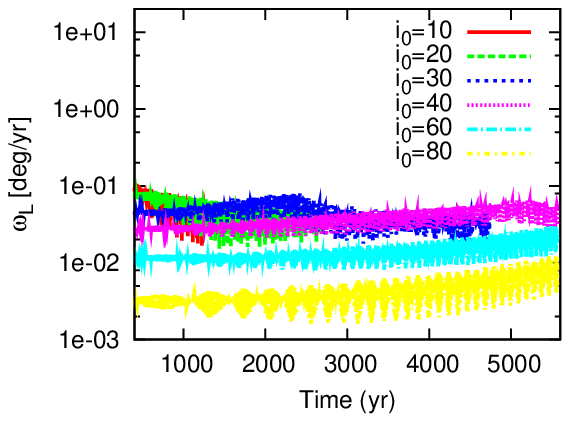}
\caption{The orbital evolution for  a $M_p=4\ \rmn{M_J}$ planet starting on an eccentric orbit with $e=0.15$ for initial relative orbital inclinations $i_0 = 0^{\circ} , 10^{\circ} , 20^{\circ} , 30^{\circ}
, 40^{\circ} , 60^{\circ}$ and $80^{\circ}$.}
\label{fig:eccentric_orbit}
\end{figure}
However, as the change in inclination of the majority of the disc 
to the initial mid plane and the corresponding  warping distortions
are relatively mild, the Lidov-Kozai mechanism applicable to orbits of high inclination
with respect to the symmetry plane of an axisymmetric potential might be expected to operate (see the discussion in Section \ref{sec:LidovK} ). As the part of 
 a Lidov-Kozai cycle, where  the orbit is nearly circular, corresponds to an extremum of the oscillation and being in the neighbourhood of a separatrix, and our
runs were typically less than a precession cycle, development of an exchange between
the inclination and eccentricity, characteristic of a Lidov-Kozai cycle,
that could take place on longer time scales,  may not
have been observed. In order to clarify this issue, we initiated runs as before, but starting with an initial orbital eccentricity of $0.15$. In that case, exchange between
the inclination and eccentricity is able to occur more rapidly such that it can be seen in our simulations
for high masses and high inclinations.
We show the results of such simulations for a planet with $M_p=4\ \rmn{M_J}$ 
starting with different initial relative orbital inclinations in Fig. \ref{fig:eccentric_orbit}.
It is seen that the evolution of the semi-major axis proceeds in almost the same manner as for the cases
that started with zero eccentricity.
However, in this case we found that although for $i_0 \le 40^{\circ},$ the eccentricity ultimately decayed,
for the cases of $i_0 = 60^{\circ}$ and $i_0 =  80^{\circ}$, significant eccentricities $ > 0.4$ developed
and were increasing at the end of the simulations. This was accompanied by a decrease in
relative  inclination in those cases that was not observed in the cases initiated with circular orbits.

Although these features are suggestive of the operation of a Lidov-Kozai like cycle for $i_0  >\sim 40^{\circ},$
 we remark that the observed decreases in inclination exceeded 
what would be expected from a conservation of the $z$ component of angular momentum
condition that would be applicable for an axisymmetric potential. 
This could be due to either the operation of dissipative decay and/or non axisymmetric effects.
For  $i_0  \le  40^{\circ},$  the decay rates of the inclinations are  about two to three times faster
compared to  those seen for simulations initiated with circular orbits.

\begin{figure}
\centering
\includegraphics[width=6cm]{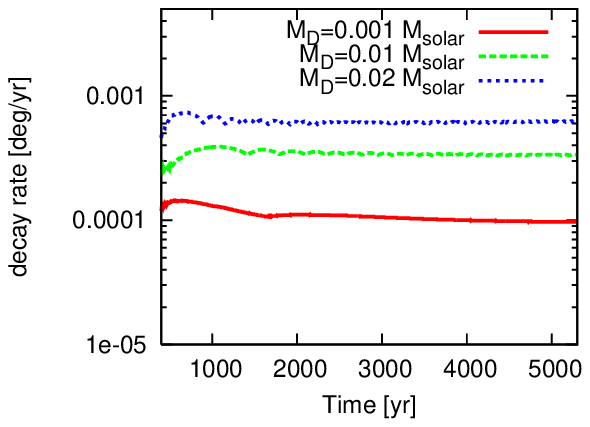}
\includegraphics[width=6cm]{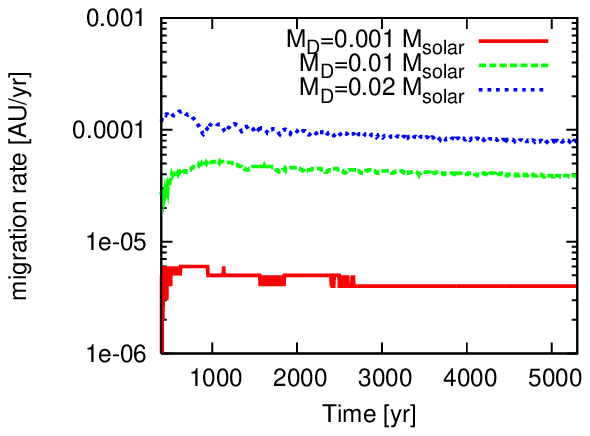}
\caption{Upper panel: the {\bf cumulative} mean of the  relative inclination  decay rate for disc masses 
of $0.001, 0.01$ and  $0.02\ \rmn{M_\odot}$. Lower panel: as above but for migration rates.}
\label{fig:rel_i_M_D}
\end{figure}

We also ran simulations  for a $6\ \rmn{M_J}$ planet starting with an initial eccentricity of $0.15$
 for  different initial relative orbital inclinations.
 Initial inclinations up to $60^\circ$ show a similar evolution to the $4\ \rmn{M_J}$ case.
 For $i_0=80^\circ$, both the inclination decrease and eccentricity increase occur more rapidly.
 After 5000 years, the relative inclination has decreased to $40^\circ$ while the eccentricity has reached 0.8.
These features are also  suggestive of the operation of a Lidov-Kozai like cycle.

The characteristic  precession period  for initial relative orbital inclinations $i_0 \leq 60^\circ$ can be
extrapolated  from Figs. \ref{fig:disc_precession_velocity} and \ref{fig:eccentric_orbit} to 
be $P_{prec} \sim 3.6\times 10^4\ \rmn{yr}.$ This  period   might also  be expected to be
comparable to the period of Kozai oscillations.  However, the warped structure of the precessing disc
may play an important role (see Figs. \ref{fig:warp_30_0} and \ref{fig:warp_40_0}) leading
to modification of the standard description.


\subsection{Dependence on the disc mass} \label{sec:M_d}
So far, all simulations have been carried out for a standard disc mass of $0.01\ \rmn·{M_\odot}$.
It is expected that the interaction between planet and disc should increase with disc mass.
Simple arguments based on dynamical friction that should be applicable to high inclination
cases indicate that decay rates should scale with the disc mass. However, as the planet mass is in general not too
different from the disc mass and 
a considerable amount of non-linearity and some distortion of the shape of the disc is present in our simulations,
such a  straightforward scaling may not be strictly valid.
Figure \ref{fig:rel_i_M_D} shows the cumulative mean of relative inclination decay rate
for $M_{p}=4\ \rmn{M_J}$ initiated on a circular orbit with $i_0=60^\circ$ for disc masses of 
$0.001, 0.01,$ and $0.02\ \rmn{M_\odot}$. The first of these being $10$ times smaller than the standard value and the last being twice as large.

It is seen that the migration rates scale as the disc mass as expected from the simple
dynamical friction argument. 
This is also the situation for the inclination decay rate when the larger of the two masses is considered.
However, the smallest disc mass has a relative inclination decay rate that is $2 - 2.5$ times faster
than expected when compared to the standard mass. A closer scrutiny of these cases indicated that the shape of the disc differed in these two cases, with different orientations of the warp.
This could result in relatively differing inclination changing torques that could be responsible for this result.

\section{Conclusions} \label{sec:concl}

We have performed SPH-simulations   in order to undertake a systematic study of the interaction
of  planets with masses  in the range $1 - 6\ \rmn{M_J}$ with initial inclinations in the range $[0;80]^\circ$ with  a circumstellar disc.
For planets initiated on a coplanar circular orbit
gap formation  occurred. We found our gap profiles to be  consistent with 
expectations from the work of  \cite{Lin1993} and \cite{Cri2006} for grid based simulations
for an adopted viscosity parameter  $\alpha_{SS}=0.025$.
Inward  migration at a  rate of $\sim 2 \times 10^4\ \rmn{AU/yr}$ for $M_p=$2, 4 and 6 $\rmn{M_J}$
was found and this is in good agreement  with previous studies and the theoretical expectation for type II migration.
For $M_p=1\ \rmn{M_J}$, which was associated with a partial gap,  the migration rate was found to be slightly slower.

  We then considered the evolution of planets initiated on the full range of  inclined circular orbits.
For these simulations, in contrast to those initiated with a finite orbital
eccentricity,  the orbits remained circular for the duration of the runs.
It is possible that phenomena such as the Lidov-Kozai effect did not have time to operate in these cases.
The inclination decay rate  was found to decrease with increasing inclination and increasing planet mass. 
This is in full accordance with expectation from estimates based on the operation of dynamical friction.
For a disc mass of $0.01M_{\odot},$ $i_0=80^\circ$ and $6\ \rmn{M_J},$ the time to decay through $10^\circ$ 
was found to be $\sim 10^{5}\ \rmn{yr}$ while for $1\ \rmn{M_J}$, it is $\sim 10^{6}\ \rmn{yr}$.
In the latter case, with the longest decay time, protoplanetary disc lifetimes are approached.
For small to intermediate $i_0$, gap formation plays a crucial role in the
determination of the inclination decay rates. Planets with larger masses are able to form gaps at higher relative inclinations.
The reduced amount of material near the planet then causes the frictional interaction to be reduced with a corresponding reduction
in the inclination decay rate.
After 200 orbits, we find the threshold initial relative inclination below which gap formation starts 
to occur to be  $i_0=10^\circ$ 
for $M_p=1\ \rmn{M_J}$, $i_0=20^\circ$ for $M_p=2\ \rmn{M_J}$, $i_0=30^\circ$ for $M_p=4\ \rmn{M_J}$
and $i_0=40^\circ$ for $M_p=6\ \rmn{M_J}$, 
For planets on highly inclined orbits, the migration rate was found to be reduced as compared to the coplanar case
due to the reduced level of interaction with the disc.
When the relative inclination approaches zero, the migration rate  approaches that of the coplanar case. 
For the case of a $4\ \rmn{M_J}$ planet initiated on a retrograde circular orbit
and different initial inclinations, we found that  the direction of 
evolution  tends towards coplanarity ($i=0^\circ$) as would be expected from interaction with the disc through
dynamical friction. The time scale for evolution decreased as $i_0$ approached  $180^\circ.$
The most extreme case was  found for $i_0=170^\circ$ where the relative
inclination  decreased by $\sim 25^\circ$ in $\sim$ 5000 years.

We also  found significant changes in the form of the disc that were produced by interaction with the  larger
planet masses. Visible warping occured in the inner parts.
The difference between the inclinations of the inner and outer part of the disc 
was found to attain up to
$10-20^\circ$. Furthermore, we calculated the total disc orientation which  can change  by up to $\sim 15^\circ$
with respect to its original mid plane.
As expected, we also found retrograde precession of both the  the total disc angular momentum vector
and the planet's angular momentum vector about the conserved angular momentum vector of the system.  This occurred at  an 
almost constant angular velocity over simulation times, with a 
 magnitude  that decreased  with increasing relative inclination. 

In addition to  simulations starting  with zero eccentricity, we initiated  a $4\ \rmn{M_J}$ planet  on  inclined orbits
 with $e=0.15$. While for $i_0 \leq 40^\circ$, the eccentricity decayed, the  cases with $i_0=60$ and $80^\circ$
 showed an eccentricity increasing for the duration of the simulations.
 The eccentricity increase was accompanied by a  relatively rapid decrease in relative inclination. 
These characteristics indicate  the possibility of  a  Lidov-Kozai  like effect operating in these cases.

In this paper, we have made a first attempt to quantify the  possible influences of  massive planets with a range  of different initial orbital  inclinations on a circumstellar disc.
Our results are reasonably consistent with theoretical estimates of evolution time scales
from dynamical friction,  although the simulated behaviour of the disc 
turns out to be  complex in some cases and was not taken into account.
In contrast to  the situation with low-mass planets,  whose influence on a disc is negligible, the results shown in this paper suggest that the interaction of a massive planet with a disc  can lead to a  strong distortion.
Furthermore the stronger frictional  interaction is likely to lead to coplanarity within disc lifetimes in most cases.
Thus highly inclined orbits are only likely to survive if they are formed after the disc has mostly dispersed.

\section*{Acknowledgements}

Xiang-Gruess acknowledges support through Leopoldina fellowship programme  (fellowship number LPDS 2009-50).
Simulations were performed using the Darwin Supercomputer of the University of Cambridge High Performance Computing Service, provided by Dell Inc. using Strategic Research Infrastructure Funding from the Higher Education Funding Council for England and funding from the Science and Technology Facilities Council.

\begin{appendix}

\section{Choice of aspect ratio in the circumplanetary disc for the modified equation of state} \label{ap:peplinski}

The main motivation of \cite{Pep2008} is to introduce a modification to the locally isothermal equation of state
in order to prevent a large amount of gas from being accreted by massive planets.
In the close vicinity of massive planets, the sound speed is  enhanced,  as  would be expected 
 for accreted optically thick material in hydrostatic
equilibrium.   This  feature is not described by a locally isothermal equation of state.
Thus the modified sound speed $c_{s,mod}$ is designed to be the sum of the unmodified soundspeed $c_{s,iso}$ and an additional positive quantity whose value increases towards the  planet and tends to zero  far from it.

\begin{figure}
\centering
\includegraphics[width=6cm]{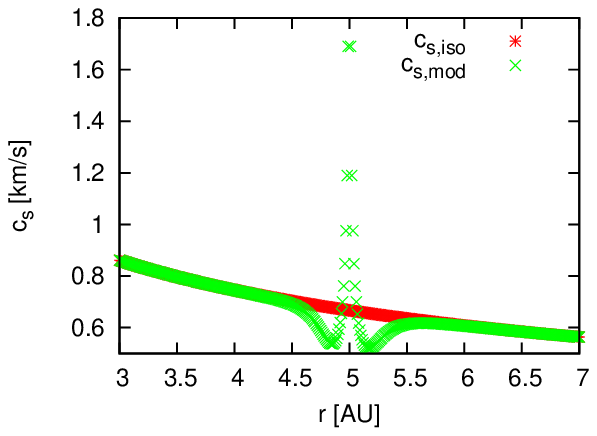} 
\includegraphics[width=6cm]{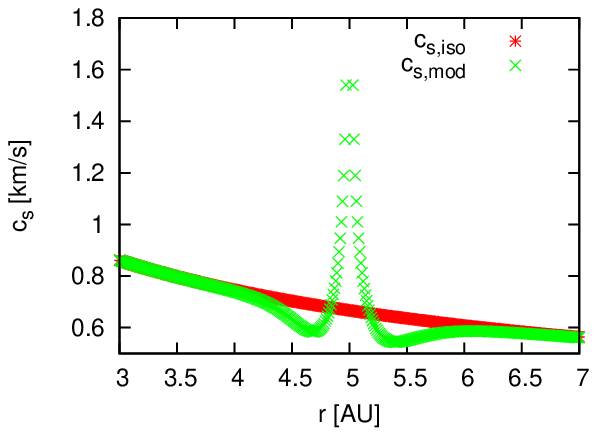} 
\includegraphics[width=6cm]{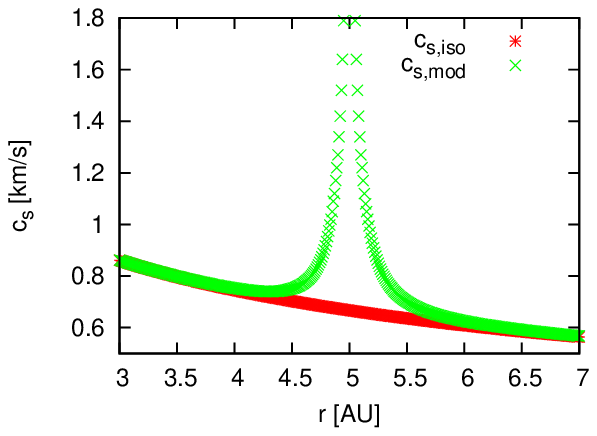} 
\caption{Modified sound speed and locally isothermal sound speed
for $M_p=0.1\ \rmn{M_J}$, $h_p=0.8$ (top); $M_p=1\ \rmn{M_J}$, $h_p=0.4$ (middle);
 $M_p=1\ \rmn{M_J}$, $h_p=0.6$ (bottom).}
\label{fig:peplinski_prob}
\end{figure}

Given the motivation indicated above, $c_{s,mod}$ should be  everywhere larger than $c_{s,iso}.$
\cite{Pep2008} have shown through various tests with  planets  with mass exceeding  $1 \rmn{M_J},$ 
 that $h_p\leq 0.4$ is appropriate in order to prevent  large amounts  of gas from being accreted.
Our studies have shown that for a  given planet mass,  $M_p$, there exists a lower limit $h_{p,lim}$
which must be exceeded  in order to guarantee that  $c_{s,mod}\geq c_{s,iso}$ globally.
 For $h_p\leq h_{p,lim}$, $c_{s,mod}\leq c_{s,iso}$  somewhere and the modification does not comply with physical requirements. 
In Table \ref{tab:c_s}, some values of $h_{p,lim}$ are listed.
\begin{table}
\centering
\begin{tabular}{|c|c|c|c|c|c|c|c|}
\hline
$M_p$ $[\rmn{M_J}]$ & 0.1 & 0.2 & 0.5 & 1 & 2 & 4 & 6 \\
\hline
$h_{p,lim}$ & 1.06 & 0.84 & 0.62 & 0.49 & 0.39 & 0.31 & 0.27 \\

\hline
\end{tabular}
\caption{Lower bounds  $h_{p,lim}$ for different planet masses }
\label{tab:c_s}
\end{table}
The lower bound  $h_{p,lim}$ decreases with increasing planet mass. For $1\ \mathrm{M_J}$, we suggest $h_p>0.49$ for the above reasons. 
Fig. \ref{fig:peplinski_prob}  illustrates two examples  where $c_{s,mod}<c_{s,iso}$ and  a  third that achieves a successful modification
of the equation of state.

\section{Ring spreading test} \label{ap:ring_test}

\begin{figure}
\centering
\includegraphics[width=7cm]{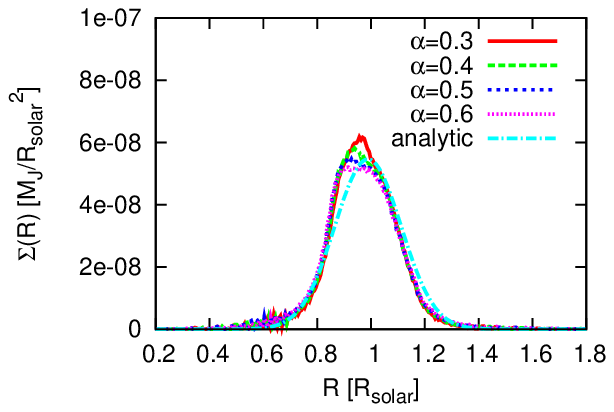}
\includegraphics[width=7cm]{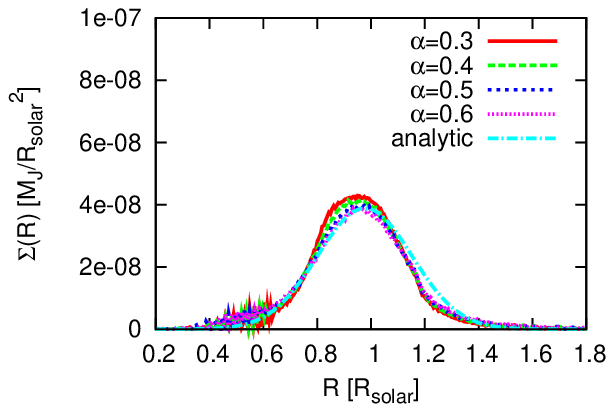}
\caption{Comparison between the analytical and numerical solutions with
 different values for the artificial viscosity parameter $\alpha$ (see Section \ref{sec:artv}) as indicated. }
\label{fig:200000}
\end{figure} 

The spreading of a viscous gas ring is often used to calibrate the effective
 viscosity \citep[e.g.][]{SpR1999}. 
The analytic solution describing an axisymmetric thin disc of gas moving around a central point mass $M_c$ under
 gravity and viscous stresses is considered.
 The vertical scale height,  $H,$ of the disc is assumed to be negligible compared to the radial length scale 
 and the viscous time scale governing  the radial inflow of mass is
 assumed to be  much larger than the dynamical time scale.
  In the thin-disc approximation, it is possible to
integrate the governing equations over the height of the disc and consider a
two dimensional flow.  Adopting  polar coordinates $(R, \varphi),$ 
there are  two velocity components $(v_R, v_\varphi)$, and the mass distribution is described by the surface density
\begin{eqnarray}
 \Sigma(R, t)= \int_{-\infty}^{+\infty} \rho dz\ , 
\end{eqnarray}
where $\rho$ is the mass density. 
 Assuming  that the radial inflow of mass is caused entirely by viscous stresses, the surface density can be shown to satisfy
 the diffusion equation \citep{Pri1981} 
\begin{eqnarray}
 \frac{\partial \Sigma}{\partial t} = \frac{3}{R} \frac{\partial}{\partial R} \left[ \sqrt{R}  \frac{\partial}{\partial R}  (\nu_0\sqrt{R} \Sigma)\right]\ . \label{eq:ring_7}
\end{eqnarray}
If the initial surface density distribution is chosen to be proportional to a $\delta$ function corresponding to
 a ring of mass $M$  localised at an initial radius $R_0,$ then the initial surface density distribution  $\Sigma_0(R)$
 is given by  
\begin{eqnarray}
 \Sigma_0(R)=\frac{M}{2 \pi R_0} \delta(R-R_0) \label{eq:ring_9}
\end{eqnarray}
The analytic solution of (\ref{eq:ring_7}) gives  $\Sigma$ at a later time as  
\begin{eqnarray}
  \Sigma(R, t)&=&\frac{M}{\pi R_0^2} \frac{1}{\tau x^{1/4}} I_{1/4} \left(\frac{2x}{\tau} \right) \exp\left( - \frac{1+x^{2}}{\tau} \right) \label{eq:ring_sigma}  \ , \\
 {\rm with}\hspace{1mm} v_R &=& \frac{6 \nu_0 }{R_0 \tau} \left[x-I_{-3/4} \left( \frac{2x}{\tau}\right) / I_{1/4} \left( \frac{2x}{\tau}\right) \right]\ .
\end{eqnarray}
Here $I$ are the modified Bessel functions. $x=R/R_0$, $\tau=12 \nu_0 t/R_0^{2}$ is the time in units of the viscous time 
scale $R_0^{2}/(12\nu_0)$ and $\nu_0$ is the kinematic viscosity \citep[see][]{Pri1981}.
We adopt the values $R_0=1\ \rmn{R_\odot}$, $M=10^{-10}\ \rmn{M_\odot}$, $M_c=1\ \rmn{M_\odot}$, $\nu_0=1.5 \cdot 10^{14}\ \rmn{cm^2/s}$ (see also \cite{SpK2003}). Since $\Sigma(t=0)$ is a $\delta$-function, the test is started by  initiating
the state variables to match conditions
 at $\tau=0.02$ and then the system is evolved forwards in time. 
The corresponding dimensionless $\alpha_{SS}$ parameter \citep{Sha1973} corresponding to this optimised value of $\nu_0$  is
\begin{eqnarray}
 \alpha_{SS}=\frac{\nu_0}{c_s H}=\frac{1.5 \cdot 10^{14}\ \rmn{cm^2/s}}{0.05^2 \sqrt{\rmn{G M_\odot R_o}}} = 0.0196 \cong 0.02\ .
\end{eqnarray}
 Results from our test calculations with  $2 \times 10^5$ SPH particles, are shown in Figure \ref{fig:200000}.  In contrast to \cite{SpK2003}, we didn't find any indications for spiral arm formation.
 In this context we note that according to these authors,  the spiral features 
 are associated with instabilities driven by shear  viscosity  and that they   used a version of SPH with a treatment of viscosity  that differs from the one we used.

\section{Resolution study} \label{ap:resolution}

\begin{figure}
\centering
\includegraphics[width=5cm]{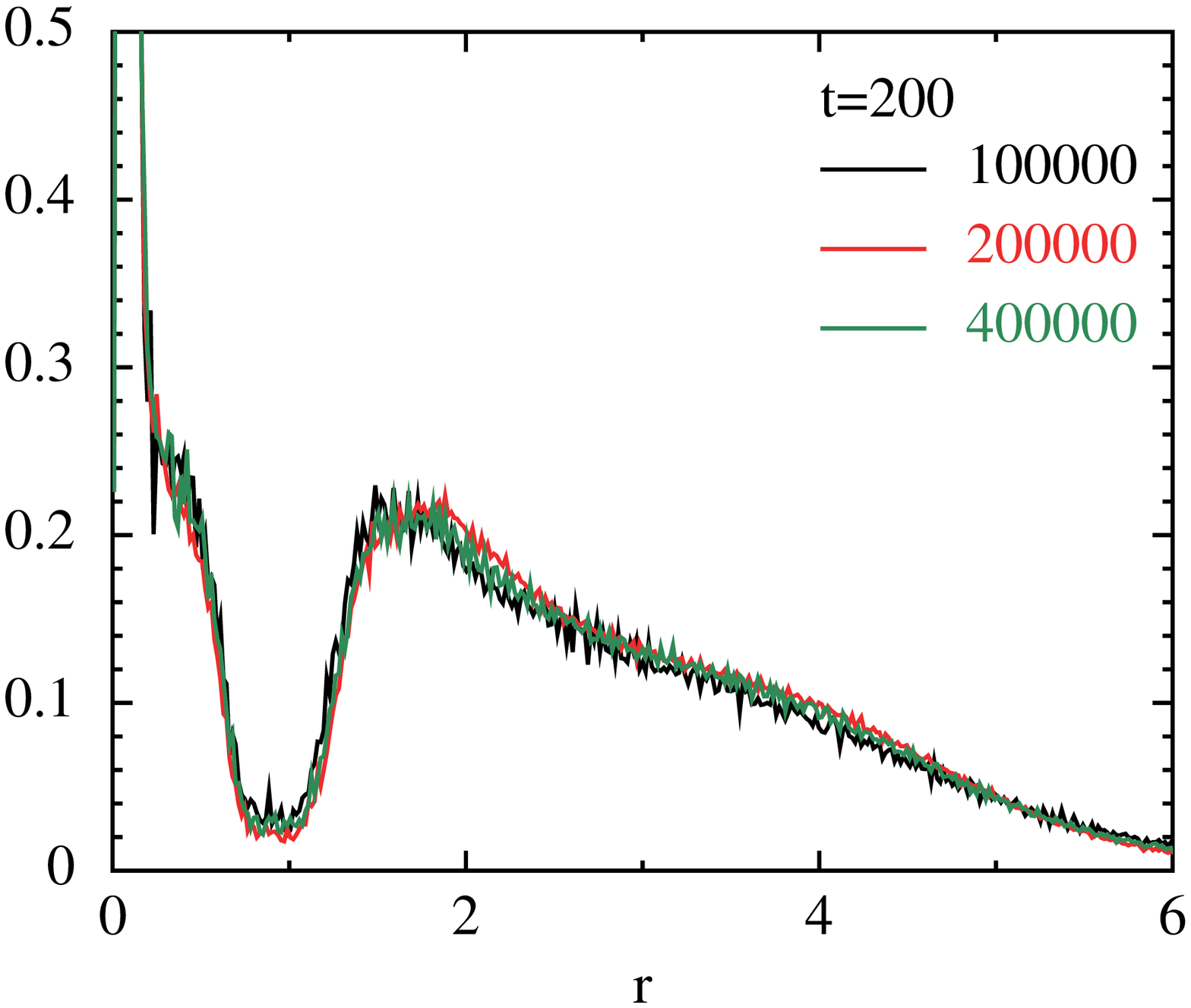} 
\includegraphics[width=6cm]{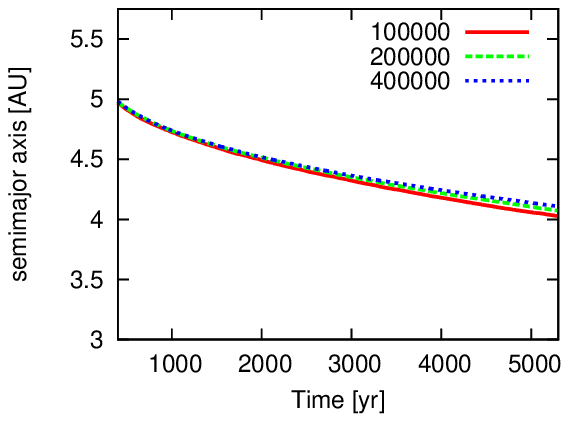}
\caption{Resolution study of  gap formation (upper panel)  and corresponding migration (lower panel)
 for  a coplanar planet with  $M_{p}=4\ \rmn{M_J}$ and a disc of $M_D=0.01\ \rmn{M_\odot}.$
 Curves corresponding to runs with different numbers of particles as indicated.}
\label{fig:resolution_coplanar}
\end{figure}
\begin{figure}
\centering
\includegraphics[width=6cm]{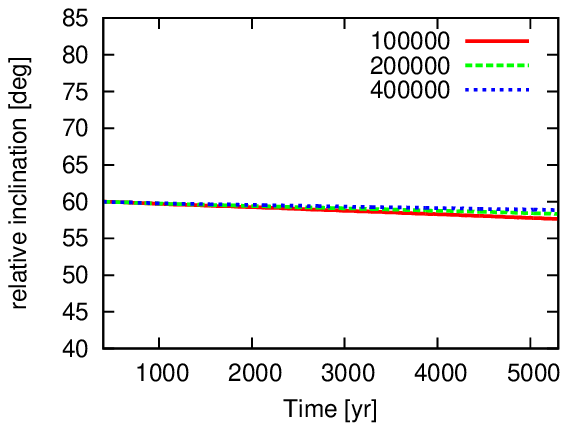}
\includegraphics[width=6cm]{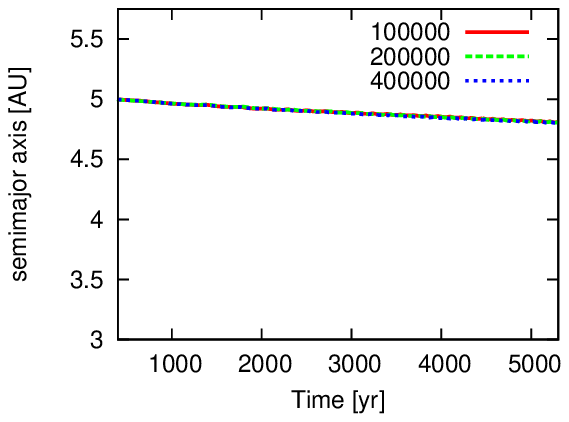}
\caption{Resolution study of relative inclination evolution  (upper panel) and  inward migration (lower panel)
for a planet  with  $M_{p}=4\ \rmn{M_J}$  and $i_0=60^\circ.$ The  disc mass was  $M_D=0.01\ \rmn{M_\odot}.$
 Curves corresponding to  runs with different numbers of particles as indicated.}
\label{fig:resolution_60}
\end{figure}

\begin{figure}
\centering
\includegraphics[width=6cm]{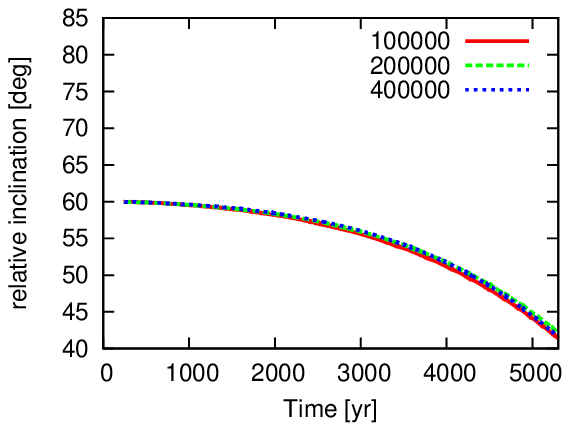}
\includegraphics[width=6cm]{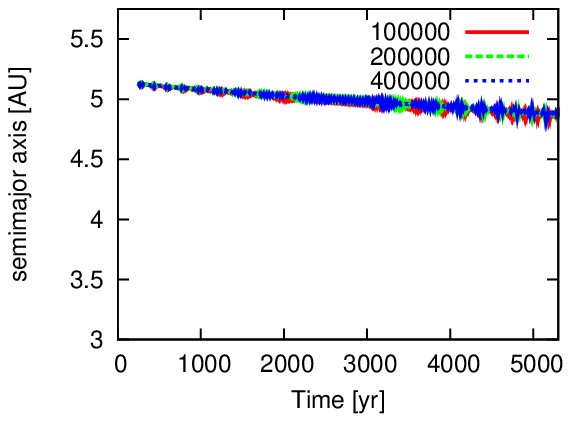}
\includegraphics[width=6cm]{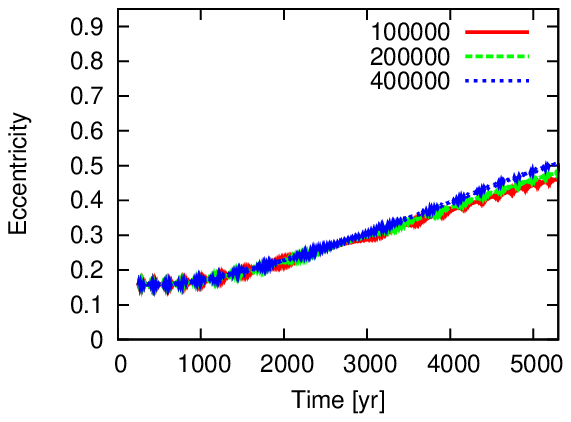}
\caption{Top two panels: as for Fig. \ref{fig:resolution_60} but starting with an initial eccentricity of 0.15.
Bottom panel: The evolution of the eccentricity is shown. }
\label{fig:resolution_60_i_ecc}
\end{figure}

In order to investigate the sensitivity of the simulations to  the number of SPH particles used,
we have performed a number of studies with differing numbers of particles
for a planet with $M_p= 4\ \rmn{M_J}$ initiated on both circular orbits,  and orbits with
an initial eccentricity,  around a central star with $M_*=1 \rmn{M_\odot}$
and interacting with a disc with $M_D=0.01 \rmn{M_\odot}.$
We considered the coplanar case, $i_0=0,$ and the  case where $i_0 = 60^\circ.$ 

Runs were performed with, $10^5, 2\times10^5,$ and $4\times 10^5$ SPH particles.
We remark that as shown by ring spreading tests, a disc modelled with SPH particles
has a non zero effective shear viscosity that is expected to decrease weakly  as the number of particles
is increased. Thus one wants to verify that a fluid disc with  an effective shear viscosity 
is appropriately modelled for the number of particles chosen.
For all these test  studies, the disc  was allowed to evolve for 30 orbits without the planet 
in order to allow any  initial inhomogenities and fluctuations to be smoothed out.
 The planet was then suddenly  introduced after 30 orbits. The sudden introduction produces transients
that rapidly decay and so produce no lasting noticeable effects.

In Fig. \ref{fig:resolution_coplanar} we show the results for initiation on a circular orbit with $i_0=0.$
Gap profiles and the evolution of the semi-major axis are plotted for the three runs with differing  numbers of SPH particles.
For both  gap profiles and the evolution of the semi-major axis, the different resolution levels
show  good agreement over the run time of the simulations. There is a hint of a decreasing shear viscosity
with increasing numbers of particles, as the semi-major axis decreases slightly less,  but this effect is small. 
 
Fig.  \ref{fig:resolution_60} shows the evolution of the relative inclination and the  semi-major axis for
the cases when the planet is initiated on a circular orbit with $i_0=60^\circ.$
The results of  simulations carried out with different numbers of particles are also  seen to be in good agreement.
In this case the evolution is expected to be driven by dynamical friction and thus expected to be  insensitive to  disc viscosity.
The evolution of the semi-major axis is indistinguishable in these cases.
The correspondence for the evolution of the relative inclination is  not quite as precise. 
 However, note that the evolution of the relative inclination is particularly sensitive because the
initial orbit is likely to be close to a separatrix and its initial evolution thus expected to be sensitive to small perturbations
(see discussion in Section \ref{sec:LidovK}).

We have also run a resolution test for a planet starting with $i_0=60^\circ$ and an initial eccentricity $e \cong 0.15$.
This study was performed to check the results described in Section \ref{sec:incecc}.
The evolution is again expected to be controlled by dynamical friction and  insensitive to disc viscosity.
The inclination and semi-major axis  evolution are  plotted in Fig. \ref{fig:resolution_60_i_ecc}.
The results of  simulations carried out with different numbers of particles are seen to be in good agreement.
The runs undertaken with different numbers of particles  thus indicate good convergence and 
that a choice of  $2\times 10^5$ SPH particles for  simulations of the type presented here is a reasonable one.

\end{appendix}

\bsp

\label{lastpage}

\end{document}